\documentclass[preprint2,trackchanges]{aastex701}
\shorttitle{SN 2025pht Progenitor Candidate}
\shortauthors{S.~D.~Van Dyk et al.}
\usepackage{hyperref}

\begin{document}

\title{The Progenitor of the Type II-Plateau SN 2025pht in NGC 1637: The Dustiest, Most Luminous Red Supergiant So Far?}

\author[0000-0001-9038-9950]{Schuyler D.~Van Dyk}
\affiliation{Caltech/IPAC, Mailcode 100-22, Pasadena, CA 91125, USA}
\email[show]{vandyk@ipac.caltech.edu}

\author[0000-0003-4610-1117]{Tam\'as Szalai}
\affiliation{Department of Experimental Physics, Institute of Physics, University of Szeged, D{\'o}m t{\'e}r 9, Szeged, 6720, Hungary}
\affiliation{MTA-ELTE Lend\"ulet "Momentum" Milky Way Research Group, Hungary}
\email{szaszi@titan.physx.u-szeged.hu}  

\author[0000-0002-5259-2314]{Gagandeep S.~Anand}
\affiliation{Space Telescope Science Institute, 3700 San Martin Drive, Baltimore, MD 21218, USA}
\email{ganand@stsci.edu}

\author[0000-0001-5955-2502]{Thomas~G.~Brink}
\affiliation{Department of Astronomy, University of California, Berkeley, CA 94720-3411, USA}
\email{tgbrink@berkeley.edu}

\author[0009-0006-5127-8290]{Noah Zimmer}
\affiliation{Department of Physics and Astronomy, Purdue University, 525 Northwestern Avenue, West Lafayette, IN 47907, USA}
\affiliation{Integrative Data Science Initiative, Purdue University, West Lafayette, IN 47907, USA}
\email{zimmer74@purdue.edu}

\author[0000-0002-0763-3885]{Dan Milisavljevic}
\affiliation{Department of Physics and Astronomy, Purdue University, 525 Northwestern Avenue, West Lafayette, IN 47907, USA}
\affiliation{Integrative Data Science Initiative, Purdue University, West Lafayette, IN 47907, USA}
\email{dmilisav@purdue.edu}

\author[0000-0003-2238-1572]{Ori D.~Fox}
\affiliation{Space Telescope Science Institute, 3700 San Martin Drive, Baltimore, MD 21218, USA}
\email{ofox@stsci.edu}

\author[0000-0001-5754-4007]{Jacob E.~Jencson}
\affiliation{Caltech/IPAC, Mailcode 100-22, Pasadena, CA 91125, USA}
\email{jjencson@ipac.caltech.edu}

\author[0000-0002-2636-6508]{WeiKang Zheng}
\affiliation{Department of Astronomy, University of California, Berkeley, CA 94720-3411, USA}
\email{weikang@berkeley.edu}

\author[0000-0003-3460-0103]{Alexei V.~Filippenko}
\affiliation{Department of Astronomy, University of California, Berkeley, CA 94720-3411, USA}
\email{afilippenko@berkeley.edu}

\author[0000-0002-9658-6151]{{\L}ukasz Wyrzykowski}
\affiliation{Astrophysics Division, National Centre for Nuclear Research, Pasteura 7, 02-093 Warsaw, Poland}
\affiliation{Astronomical Observatory, University of Warsaw, Al.~Ujazdowskie~4, 00-478 Warsaw, Poland}
\email{lw@astrouw.edu.pl}

\author[0000-0001-8916-8050]{Przemys{\l}aw J.~Miko{\l}ajczyk}
\affiliation{Astronomical Institute, University of Wroc{\l}aw, ul.~Miko{\l}aja~Kopernika~11, 51-622 Wroc{\l}aw, Poland}
\affiliation{Astrophysics Division, National Centre for Nuclear Research, Pasteura 7, 02-093 Warsaw, Poland}
\email{mikolajczyk@astro.uni.wroc.pl}

\author[0000-0003-4960-7463]{Krzysztof Kotysz}
\affiliation{Astronomical Institute, University of Wroc{\l}aw, ul.~Miko{\l}aja~Kopernika~11, 51-622 Wroc{\l}aw, Poland}
\affiliation{Astronomical Observatory, University of Warsaw, Al.~Ujazdowskie~4, 00-478 Warsaw, Poland}
\email{kotysz@astro.uni.wroc.pl}

\author[0000-0003-0125-8700]{F.-J.~Hambsch}
\affiliation{Vereniging Voor Sterrenkunde (VVS), Zeeweg 96, B-8200 Brugge, Belgium}
\affiliation{Bundesdeutsche Arbeitsgemeinschaft für Veränderliche Sterne (BAV), Munsterdamm 90, 12169 Berlin, Germany}
\email{hambsch@telenet.be}

\author[]{Agata Wo\'{z}niak}
\affiliation{Janusz Gil Institute of Astronomy, University of Zielona Gora, Szafrana 2, 65-516 Zielona Gora, Poland}
\email{agatawozniak2304@gmail.com}

\author[0000-0001-5836-9503]{Micha{\l} \.{Z}ejmo}
\affiliation{Janusz Gil Institute of Astronomy, University of Zielona Gora, Szafrana 2, 65-516 Zielona Gora, Poland}
\email{m.zejmo@ia.uz.zgora.pl}

\author[0000-0003-4254-2724]{Andrea Reguitti}
\affiliation{INAF – Osservatorio Astronomico di Padova, Vicolo dell'Osservatorio 5, I-35122 Padova, Italy}
\affiliation{INAF – Osservatorio Astronomico di Brera, Via E. Bianchi 46, 23807 Merate (LC), Italy}
\email{andrea.reguitti@inaf.it}

\begin{abstract}
We provide a characterization of the red supergiant (RSG) progenitor candidate for the nearby Type II-plateau supernova (SN) 2025pht in NGC 1637. The star was first detectable in 2001 by the {\sl Hubble Space Telescope\/} ({\sl HST}) and then again in a dozen bands by the {\sl James Webb Space Telescope\/} ({\sl JWST}) in 2024. This ``quasi-snapshot'' of the star's nature almost immediately prior to explosion is unprecedented. The RSG varied in brightness, and we posit that it could have been a pulsating variable, possibly with a long period of $\sim 660$ days. The largest uncertainty is the host-galaxy distance, which we establish to be $11.67\pm0.27$ Mpc. The star was also heavily extinguished by interstellar dust internal to the host, with visual extinction $A_V({\rm host})\approx1.7$ mag (total $A_V({\rm tot})\approx1.8$ mag). Dust radiative-transfer modeling reveals the star's circumstellar medium to be quite dusty and silicate-rich, yielding a bolometric luminosity as high as $\log(L_{\rm bol}/L_{\odot})=5.16\pm 0.03$ and a cool effective temperature $T_{\rm eff}=2100$--2500 K. The available {\sl HST\/} optical data had no bearing on the shape of the candidate's observed spectral energy distribution --- for the first time, without the archival {\sl JWST\/} observations we would not have been able to detect and characterize the candidate at all. The SN 2025pht progenitor candidate, although  similar to that of SN 2023ixf, may be the most luminous star identified to date.
%250 word limit for the abstract. 
\end{abstract}


\section{Introduction}

Core-collapse supernovae (SNe) are thought to arise from stars with initial masses $M_{\rm ini}\gtrsim 8\ M_{\odot}$. Single stars in this mass range are further thought to reach the end of their lives in the red supergiant (RSG) phase. The SNe that result from the explosions of such stars, with massive extended hydrogen-rich envelopes, should be observed as the H-rich Type II SNe, specifically, the SNe II-Plateau, or II-P. Indeed, strong observational evidence exists that SN II-P progenitors are RSGs, specifically through direct identification and characterization of the progenitor stars in pre-explosion images \citep[e.g., summaries by][]{Smartt2009,Smartt2015,VanDyk2017}.

\citet{VanDyk2025} recently reviewed RSGs as SN progenitors. The conclusions, based on the cases to date, were that low-luminosity SNe II-P arose from progenitor RSGs in a relatively low range of bolometric luminosity $\log (L_{\rm bol}/L_{\odot}) \approx 4$--4.7 \citep[see also, e.g.,][]{Lisakov2018,Rodriguez2022}, and that these stars can generally be described into the near-infrared by a bare stellar atmosphere model, typically of effective temperature $T_{\rm eff}\approx 3400$--3700~K \citep[see also, e.g.,][]{Levesque2005,Levesque2017}. Likely all RSGs at the pre-SN stage are embedded to some extent in circumstellar material (CSM), established via mass loss from the star. For RSGs with $\log (L_{\rm bol}/L_{\odot}) \gtrsim 4.7$ the CSM may become more massive and dense, and therefore of a higher dust content, as evinced, for instance, by the characterization and modeling of Galactic RSGs \citep[e.g.,][]{Massey2005,Verhoelst2009}. Recent higher-luminosity SN II-P events, such as SN 2012aw \citep{VanDyk2012,Kochanek2012,Fraser2012}, SN 2017eaw \citep{Kilpatrick2018,VanDyk2019,Rui2019}, and SN 2023ixf \citep[e.g.,][]{Kilpatrick2023,Jencson2023,Xiang2024,VanDyk2024}, have exemplified the premise that some SNe may arise from higher-luminosity, dusty RSGs. Moreover, SN II-P progenitors appear to be limited to stars with $\log (L_{\rm bol}/L_{\odot}) \lesssim 5.2$, although the limit to known RSG luminosities appears to be $\log (L_{\rm bol}/L_{\odot}) \approx 5.7$ \citep{Humphreys1979,Humphreys2025}, leading to the so-called ``RSG problem'' \citep[e.g.,][]{Smartt2009,Walmswell2012,Davies2020,Kochanek2020,Beasor2025}.

Many known RSGs in Local Group host galaxies are semiregular, pulsational, long-period variables (LPVs, with periods of hundreds to thousands of days; \citealt{Kiss2006,Yang2011,Yang2012,Soraisam2018,Ren2019}). In fact, the progenitor candidate of SN 2023ixf in Messier 101 was convincingly shown to be a variable with fundamental period $P \approx1000$--1100 days \citep{Jencson2023,Soraisam2023}. 
The particular subtype of SN~II could possibly be a manifestation of the phase of the light curve, and the extent of the stellar envelope, at which the explosion occurs \citep{Bronner2025}.
Mass loss is very important for the conditions in which the SN develops \citep[e.g.,][]{Dessart2017}. Most SN shock-CSM interaction can be explained by an extensive, dense atmosphere from which the rate of stellar mass loss is generally fairly low ($\dot M< 10^{-4}\ M_{\odot}\ {\rm yr}^{-1}$ for stars with $M_{\rm ini}\lesssim 20\ M_{\odot}$; \citealt{vanLoon2025}).
However, pulsations in RSGs, especially, nonradial pulsations and overtone modes, could drive instabilities at late evolutionary stages, leading to enhanced mass loss and ejections \citep{Suzuki2025,Sengupta2026}, possibly even via a ``superwind'' \citep{Yoon2010}, altering the CSM that the SN shock encounters at explosion \citep{Goldberg2020}.
As the star loses mass before explosion, the luminosity-to-mass ratio will increase, further driving pulsational amplitudes \citep{Laplace2026}.
Abrupt, massive outbursts from the star $\lesssim 1$ yr before its demise could also account for the optical appearance of the resulting SN \citep{Davies2022}, specifically at or near shock breakout \citep{Dessart2026}.

Here we present a characterization of the progenitor of SN 2025pht in the nearby spiral host galaxy NGC 1637. \citet{Stanek2025} reported its discovery as a new transient on 2025 June 29 (UTC dates are used throughout this paper), and \citet{Strader2025} later classified it from an optical spectrum obtained on July 3 as SN II-P. Little else has been reported on the SN, other than that covered by \citet{Kilpatrick2025}. D.~Hodges added, also on July 3, a comment to the Transient Name Server online page for the SN\footnote{https://www.wis-tns.org/object/2025pht} that, based on its position reported by \citet{Stanek2025}, the SN was very near an apparent blue supergiant, which was presented as a possible progenitor candidate. Subsequently, on July 16, \citet{Perez2025} reported a position for the SN measured from ground-based Las Cumbres Observatory imaging, RA= $04^{\rm hr}41^{\rm m}28{\fs}872$, Dec = $-02\arcdeg51^{\prime}55{\farcs}84$ (J2000; $\pm0{\farcs}08$), that is $0{\farcs}57$ from the \citet{Stanek2025} position. The new position corresponded to a source detected in images obtained by the {\sl James Webb Space Telescope\/} ({\sl JWST}), which made it potentially more likely to be a luminous red star and plausibly a better progenitor candidate, given the SN type, than the one suggested by Hodges. 

We present our analysis of this progenitor candidate, based on the broad assumption that the \citet{Perez2025} identification is correct. (\citealt{Kilpatrick2025} have subsequently demonstrated via precise relative astrometry that this star is indeed the best progenitor candidate.) We will show that the star itself was behind a substantial curtain of internal host extinction. What we find, based on the pre-SN observational evidence, is that the star is highly likely to have been a luminous RSG, with a significant amount of CSM dust. 

We note that the host galaxy of SN 2025pht was also the host of the well-studied SN II-P 1999em \citep[e.g.,][]{Hamuy2001,Leonard2002,Elmhamdi2003,Utrobin2007}. The progenitor of that SN was not detected in pre-SN ground-based optical imaging \citep{Smartt2002}. 
The inclination for the host galaxy, of type SBc, is $i\approx39\degr$ (from axial ratio $b/a=0.78$; position angle $PA=33\degr$; \citealt{Jarrett2003}, via the NASA/IPAC Extragalactic Database, NED\footnote{https://ned.ipac.caltech.edu/}). The host is a rather small spiral, with, respectively, major and minor axis (isophotal) diameters of $153{\farcs}2$ and $119{\farcs}5$ \citep{Jarrett2003}, which are $\sim 8.0$ and $\sim 6.2$ kpc, respectively, for the distance we adopt below. We also adopted a host-galaxy redshift $z=0.002392$, via NED.

This paper is organized as follows. The available observations of the SN and its site are summarized in Section~\ref{sec:observations}; analysis of the data, including the properties of the SN, photometry of the SN progenitor candidate, the possible variability of the candidate, the reddening to the SN, and the distance to the host galaxy, is presented in Section~\ref{sec:analysis}; the subsequent inferred properties of the progenitor candidate are put forward in Section~\ref{sec:progenitor}; and, finally, a discussion and conclusions are offered in Section~\ref{sec:discussion}.

\section{Observations} \label{sec:observations}

The SN 2025pht site was serendipitously captured prior to explosion by a number of space-based facilities. Here we concentrate on the most salient data sources, which, as \citet{Perez2025} pointed out, are from the {\sl Hubble Space Telescope\/} ({\sl HST}) and {\sl JWST}. See Table~\ref{tab:observations} for an inventory. 

\begin{deluxetable}{ccccccc}
%\digitalasset
\tablewidth{0pt}
\tablecaption{Inventory of Observations\label{tab:observations}}
\tablehead{
\colhead{Instrument} & \colhead{Band} & \colhead{Observation} & \colhead{space{\textunderscore}phot} & \colhead{Dolphot} & \colhead{Flux density $f_{\nu}$} \\
\colhead{} & \colhead{} & \colhead{Date} & \colhead{mag (AB)} & \colhead{mag (AB)} & \colhead{($\mu$Jy)}
}
\startdata
WFPC2     & F450W                  & 2001 Aug 12 & \nodata         & $<24.9$         & $<0.442$\\
          & F555W\tablenotemark{a} &             & \nodata         & \nodata         &  \nodata\\
          & F606W                  & 1994 Sep 11 & \nodata         & $<24.7$         & $<0.532$\\
          & F814W\tablenotemark{a} &             & \nodata         & \nodata         &  \nodata\\
WFC3/UVIS & F275W                  & 2024 Aug  3 & \nodata         & $<26.6$         & $<0.094$\\
          & F336W                  &             & \nodata         & $<26.6$         & $<0.096$\\
          & F438W                  &             & \nodata         & $<26.4$         & $<0.109$\\
          & F555W                  &             & \nodata         & $<26.6$         & $<0.096$\\
          & F657N                  &             & \nodata         & $<25.1$         & $<0.360$\\
          & F814W                  &             & \nodata         & $<26.0$         & $<0.158$\\
NIRCam    & F150W                  & 2024 Feb 5  & $22.23(0.17)$   & $22.184(0.013)$ & $4.64(0.71)$\\ 
          &                        & 2024 Oct 8  & $22.30(0.05)$   & $22.214(0.004)$ & $4.37(0.20)$\\
          & F164N                  & 2024 Oct 8  & $21.76(0.06)$   & $21.682(0.012)$ & $7.18(0.40)$\\
          & F187N                  & 2024 Feb 5  & $23.08(2.02)$\tablenotemark{b} & $21.895(0.060)$ &  $2.1(+3.9,-2.1)$\\ 
          &                        & 2024 Oct 8  & $22.11(0.27)$   & $21.947(0.014)$ &  $5.22(1.30)$\\
          & F200W                  & 2024 Oct 8  & $21.70(0.06)$   & $21.613(0.003)$ &  $7.59(0.44)$\\
          & F212N                  &             & $21.41(0.20)$   & $21.275(0.009)$ &  $9.88(1.80)$\\
          & F277W                  &             & $21.48(0.03)$   & $21.401(0.002)$ &  $9.29(0.26)$\\
          & F300M                  & 2024 Feb 5  & $21.17(0.07)$   & $21.124(0.008)$ &  $12.31(0.77)$\\
          & F335M                  &             & $20.97(0.05)$   & $20.939(0.004)$ &  $14.80(0.71)$\\ 
          &                        & 2024 Oct 8  & $21.02(0.03)$   & $20.982(0.002)$ & $14.17(0.45)$\\
          & F360M                  &             & $20.92(0.03)$   & $20.874(0.002)$ & $15.57(0.37)$\\
          & F405N                  &             & $20.76(0.09)$   & $20.807(0.007)$ & $18.09(1.48)$\\
          & F444W                  &             & $20.91(0.03)$   & $20.858(0.002)$ & $15.64(0.50)$\\
MIRI      & F770W                  & 2024 Feb 5  & $20.93(0.19)$   & $20.506(0.010)$ & $15.4(2.7)$\\
          & F2100W                 &             & $<20.7$           & \nodata & $<18.3$\\
\enddata
\tablenotetext{a}{See Table~\ref{tab:light_curve}.}
\tablenotetext{b}{Only two of the four individual ``{\em {\textunderscore}nrcb3{\textunderscore}cal} '' frame files were available, leading to an imprecise measurement. In this case, we chose to leave this value out of an uncertainty-weighted mean and only considered the October 8 measurement.}
\end{deluxetable}

The {\sl HST\/} observations include those from 2001 August 12 with the Wide Field and Planetary Camera 2 (WFPC2; \citealt{Holtzman1995}) instrument in bands F450W (total exposure time, 460 s) and F814W (also 460 s) by program GO-9042 (PI S.~Smartt). Further WFPC2 observations include a dedicated  campaign observing the host galaxy by GO-9155 (PI D.~Leonard) in F555W and F814W on several epochs between 2001 September 2 and October 31 (1100 s per exposure in each band); see Table~\ref{tab:light_curve}. Additionally, the host was observed by GO-5446 (PI G.~Illingworth) in F606W (160 s) in 1994 September. The SN site was not located in the observations by GO-9041 (PI S.~Smartt).

\begin{deluxetable}{cccc}
%\digitalasset
\tablewidth{0pt}
\tablecaption{{\sl HST\/} WFPC2 F555W and F814W Data\tablenotemark{a}\label{tab:light_curve}}
\tablehead{
\nocolhead{Obs.~Date} & \colhead{MJD} & \colhead{F555W Mag\tablenotemark{b}} & \colhead{F814W Mag}\\
\colhead{} & \colhead{} & \colhead{(Vega)} & \colhead{(Vega)}
}
\startdata
2001 Aug 12 & 52133.34 & \nodata & $24.95 \pm 0.58$\\
2001 Sep 02 & 52154.20 & $<26.3$ & $24.11 \pm 0.10$\\
2001 Sep 10 & 52162.96 & $<26.4$ & $23.91 \pm 0.09$\\
2001 Sep 20 & 52172.99 & $<26.4$ & \nodata\\
2001 Sep 23 & 52175.52 & $<25.9$ & \nodata\\
2001 Sep 26 & 52178.54 & $<26.5$ & $23.86 \pm 0.08$\\
2001 Sep 29 & 52181.88 & $<26.4$ & $23.92 \pm 0.09$\\
2001 Oct 03 & 52185.55 & $<26.5$ & \nodata\\
2001 Oct 07 & 52189.29 & $<26.4$ & \nodata\\
2001 Oct 12 & 52194.64 & $<26.4$ & \nodata\\
2001 Oct 18 & 52200.66 & $<26.5$ & $24.03 \pm 0.10$\\
2001 Oct 24 & 52206.61 & $<26.4$ & \nodata\\
2001 Oct 31 & 52213.63 & $<26.0$ & $23.76 \pm 0.08$\\
\enddata
\tablenotetext{a}{From programs GO-9042 (PI S.~Smartt) and GO-9155 (PI D.~Leonard).}
\tablenotetext{b}{Upper limits are 5$\sigma$ from {\tt Dolphot}; see \citet{VanDyk2023}.}
\end{deluxetable}

Observations of the host with the Wide Field Camera 3 (WFC3; \citealt{Pagul2024}) in the UVIS channel were conducted in 2024 August by GO-17502 (PI D.~Thilker) in a number of bands: F275W (2075 s), F336W (1000 s), F438W (1050 s), F555W (625~s), F657N (2020 s), and F814W (749 s).

The available {\sl JWST\/} observations are from two different programs, GO-3707 (PI A.~Leroy) on 2024 February 5 and GO-4793 (PI E.~Schinnerer) on 2024 October 8. Note that these epochs correspond to only 510 and 264 days (respectively) before discovery of the SN. The observations by the latter program were only in the NIRCam instrument \citep{Rieke2023}, at F150W (859 s), F164N (1632 s), F187N (1632 s), F200W (859 s), F212N (1632 s), F277W (859 s), F335M (1632 s), F360M (1632 s), F405N (1632 s), and F444W (859 s). For the former program NIRCam was also used, at F150W (215 s), F187N (387 s), F300M (215 s), and F335M (387 s), as well as with the MIRI instrument \citep{Wright2023,Dicken2024} at F770W (178 s) and F2100W (688 s).

\begin{figure}
\plotone{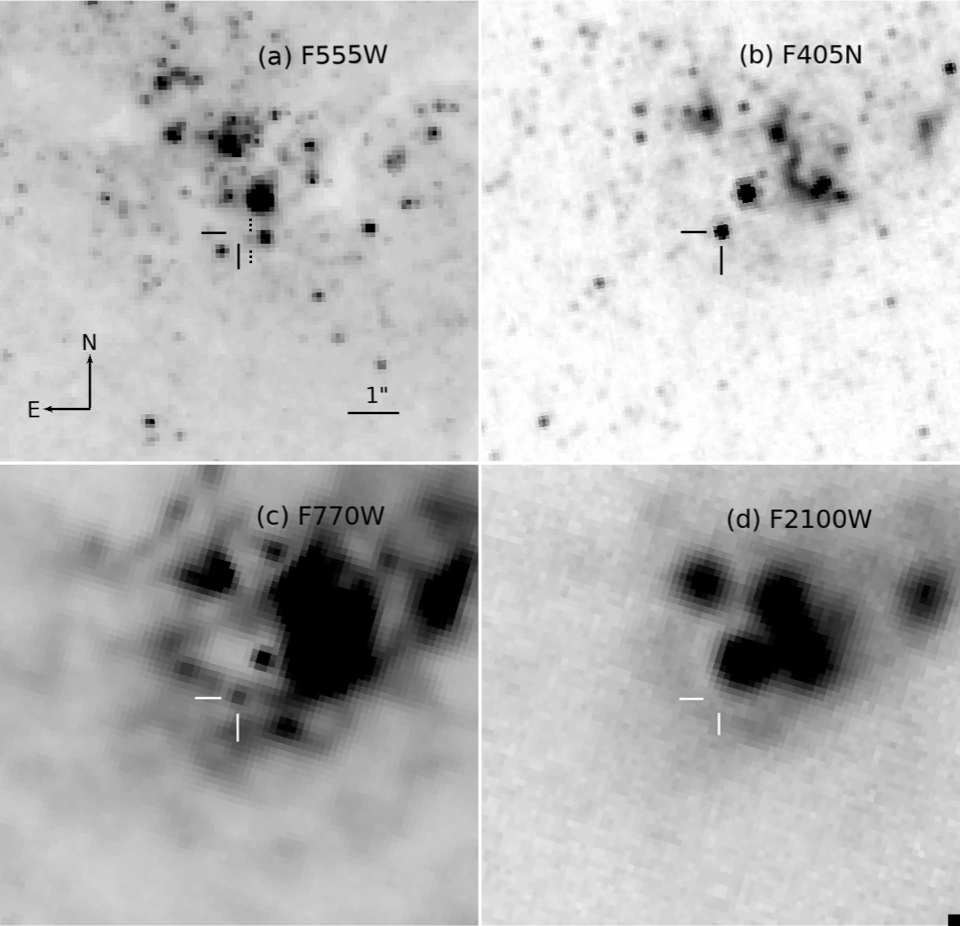}
\caption{Portions of images in which the SN 2025pht site was serendipitously captured, all shown to the same orientation and scale. The site location corresponds to that of the progenitor candidate identified by \citet{Perez2025} and indicated by solid tickmarks in each panel. Panel (a) is a coaddition of all the {\sl HST\/} WFPC2 F555W images obtained of the host galaxy from 2001 September 2 to October 31; panel (b) is a {\sl JWST \/} NIRCam F405N image from 2024 October 8; panel (c) is a {\sl JWST \/} MIRI F770W image from 2024 February 5; and, panel (d) is a MIRI image from the same date in F2100W. The candidate is well detected in both F405N and F770W, but not detected in either F555W or F2100W. North is up and east is to the left in all four panels. See also \citet{Kilpatrick2025}.}
\label{fig:montage}
\end{figure}

We show the SN site in a sampling of the {\sl HST\/} and {\sl JWST\/} data in Figure~\ref{fig:montage}. (\citealt{Kilpatrick2025} had already presented the data in F336W, F814W, F150W, F277W, and F444W; here we repeat showing F555W and F770W.) In all {\sl JWST\/} bands, except for the MIRI F2100W, the progenitor candidate is detected brightly and distinctly. In most bands the candidate is relatively isolated from its neighboring stars and nebular emission. The star is $\sim 0{\farcs}8$ southeast of what appears, by its brightness and color across both {\sl HST\/} and {\sl JWST\/} bands, to be a compact stellar cluster. At F770W, the MIRI band sensitive to emission from the 7.7 $\mu$m polycyclic aromatic hydrocarbon (PAH) feature, the star appears to be located along a ridge of a large, interstellar bubble in which the compact cluster is located.

The progenitor candidate is not detected in any of the {\sl HST\/} bands, except for the F814W observations in 2001 (see Table~\ref{tab:light_curve}). We surmise that the faint object immediately to the southwest of the SN location, as seen in the F555W data from 2001 (Figure~\ref{fig:montage} (a); indicated by dotted tickmarks), that \citet{Perez2025} had associated with the progenitor candidate, is actually unrelated to the star, since it does not correspond to the candidate's position in the F814W images from the same date range.

The {\sl Spitzer Space Telescope\/} also captured the SN site with the Infrared Array Camera (IRAC, \citealt{Fazio2004}; both during the cryogenic and post-cryogenic missions), the Multiband Imaging Photometer for Spitzer (MIPS; \citealt{Rieke2004}), and the Peak-up Imaging (PUI) mode of Infrared Spectrograph (IRS; \citealt{Houck2004}; only data with the Blue channel contained the site). The IRAC observations were a single epoch (2007 October~18) of SN 1999em by program id (PID) 40010 (PI M.~Meixner), as well as by PID 12084 (PI K.~Krafton; post-cryo) on 2016 May 10. Additional post-cryo observations of the host galaxy were conducted by PID 61068 (PI K.~Sheth) on 2011. The MIPS and PUI observations were also conducted by PID 40010 on 2007 September 25 and October 12, respectively. However, unlike for the recent case of SN 2023ixf \citep[e.g.,][]{Kilpatrick2023,Jencson2023,Soraisam2023}, these data are not useful, since the SN site is spatially confused with, assumedly, the compact star cluster and overall star-forming regions to the northwest.

Other facilities also observed the field, such as {\sl WISE\/} \citep{Wright2010} and {\sl NEOWISE\/} \citep[e.g.,][]{Mainzer2011}; however, none of the available data have sufficiently adequate  spatial resolution. Together with {\sl Spitzer}, these resources are far less constraining than {\sl HST\/} and {\sl JWST\/} and therefore have lower value. Thus, we dispense with further discussion of all of these observations.

Detailed multiband, optical light-curve data for SN 2025pht\footnote{https://bhtom.space/public/targets/SN2025pht}  were obtained by the BHTOM.space network \citep{BHTOM1,BHTOM2}, as  \citet{Mikolajczyk2025} reported.

Finally, we conducted optical spectroscopy of SN 2025pht with the Kast double spectrograph on the Shane 3 m telescope at Lick Observatory, California. We obtained spectra at three different epochs, 2025 July 31.50, August 21.51, and August 30.50, with the 2\arcsec-wide slit, the D57 dichroic, the 600/4310 grism, and the 300/7500 grating. This configuration has a combined wavelength range of $\sim 3600$--10,700 \AA\ and a spectral resolving power $ R \approx 800$. The slit was oriented at or near the parallactic angle, in order to minimize slit losses caused by atmospheric dispersion \citep{Filippenko1982}.

Data were reduced following standard techniques for CCD processing and spectrum extraction based on standard {\tt IRAF} routines \citep{Tody1986} and custom Python and {\tt IDL} software (\citealt{Silverman2012}; see https://github.com/ishivvers/TheKastShiv). Low-order polynomial fits to comparison-lamp spectra were established to calibrate the wavelength scale, and small adjustments attained from night-sky lines in the target frames were applied. Spectra were flux-calibrated via observations of spectrophotometric standard stars observed on the same nights at similar airmasses, and in an identical instrumental configuration.

All of the {\sl HST\/} and {\sl JWST\/} data used in this paper can be found in MAST: \dataset[10.17909/esvy-9c86]{http://dx.doi.org/10.17909/esvy-9c86}.

\section{Analysis}\label{sec:analysis}

\subsection{Properties of the Supernova}\label{sec:supernova}

\begin{figure}
\plotone{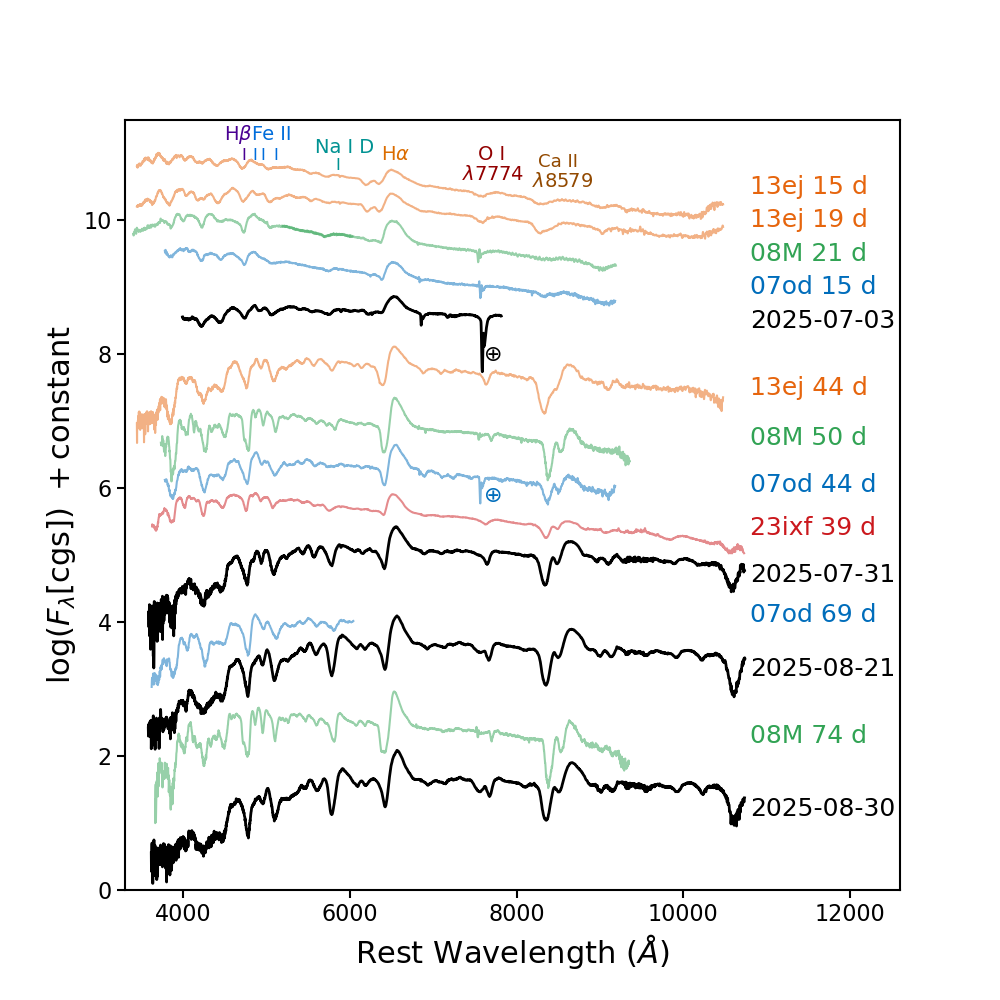}
\caption{Optical spectra of SN 2025pht from 2025 July 3 \citep{Strader2025}, and from July 31, August 21, and August 30 (this study). Various spectral features, including telluric absorption, are indicated. Also shown for comparison are spectra at various ages of the SNe II-P SN 2007od and SN 2008M \citep{Gutierrez2017}, SN 2013ej \citep{Dhungana2016}, and SN 2023ixf \citep{Zheng2025}. All spectra have been corrected for the redshifts of their host galaxies based on the values in NED. None of the spectra shown has been corrected for reddening.}
\label{fig:spectracomp}
\end{figure}

The spectra of SN 2025pht, which include both the \citealt{Strader2025} classification spectrum and the Kast spectra that we have obtained here, are shown in Figure~\ref{fig:spectracomp}. Via, for instance, {\tt GELATO} \citep{Harutyunyan2008}, the highest quality of fits to the July 3 classification spectrum were with the Type II-P SN 2009bw \citep{Inserra2012} at 18--20 days and SN 2007od \citep{Andrews2010,Inserra2011} at 14 days. Again, for the July 31 spectrum, among the highest fits were, once again, SN 2009bw at 37--39 days and SN 2007od at 46--59 days. 

For comparison to SN 2025pht we therefore show spectra of SN 2007od \citep{Gutierrez2017} in the figure. Additionally, we considered spectra of the SNe~II-P SN 2008M (\citealt{Gutierrez2017}), SN 2013ej \citep{Dhungana2016}, and the recent SN 2023ixf (\citealt{Zheng2025}; see also, e.g., \citealt{Li2025a}). We obtained the SN 2013ej spectra via WISeREP\footnote{https://www.wiserep.org/} \citep{Yaron2012}. (Unfortunately, to our knowledge, spectra of SN 2009bw have not been publicly archived.)  All of the spectra have been corrected for the redshifts of their host galaxies based on the values in NED.

The SN 2025pht spectra  resemble those of SN 2013ej, and similarities also exist with SN 2007od, although less so with SN 2008M. SN 2025pht appears to have less overall resemblance to SN 2023ixf, so we preclude any further spectral comparisons. Interestingly, both SN 2007od \citep{Andrews2010,Inserra2011} and SN 2013ej \citep{Huang2015,Chakraborti2016,Mauerhan2017} have shown evidence for CSM interaction, although it was likely stronger for the former than the latter.

From the spectral comparison we determined that the classification spectrum from July 3 \citep{Strader2025} is consistent with being from day $\approx 15$--21, similar to the results from {\tt GELATO}; if day 15, then the explosion epoch, $t_0$, for SN 2025pht would be MJD $\approx60844.4$ (2025 June 18.4), and if day 21 (2025 June 12.4), then $t_0\approx60838.4$. Given the uncertainty in the age, we adopt the average, $t_0\approx60841.4$ (2025 June 15.4), $\pm \sim3$ days. In a similar vein, the spectrum from July 31 appears to be from $\sim 44$ days, which would be consistent with age $\sim 46$ days based on the assumed date of explosion. The August 21 spectrum would represent an age of $\sim 67$ days, and the August 30 spectrum $\sim 76$ days. 

\begin{figure}
\plotone{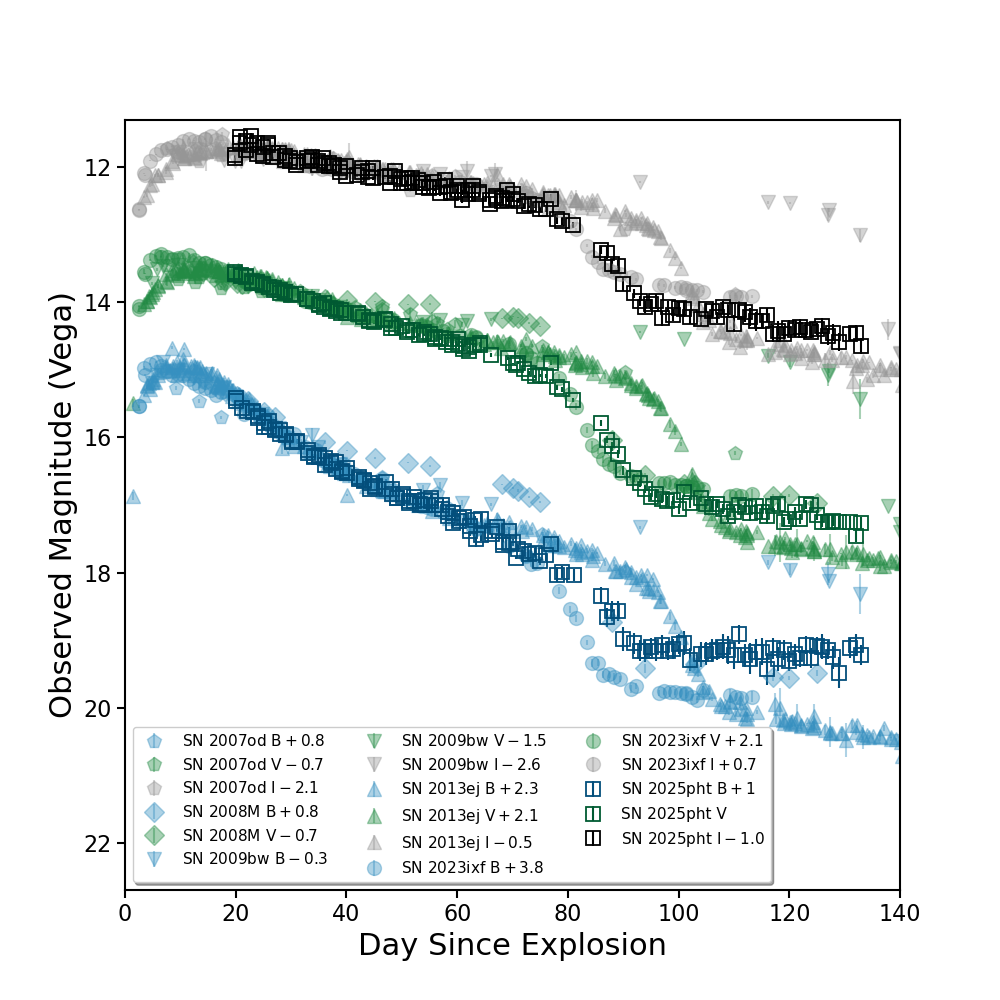}
\caption{Optical $BVI$ (Vega) light curves of SN 2025pht from BHTOM \citep{Mikolajczyk2025}. For comparison we show light curves in these bands for SN 2007od \citep{Inserra2011,Anderson2024}, SN 2008M \citep{Anderson2024}, SN 2009bw \citep{Inserra2012}, and SN 2013ej \citep{Valenti2014,Bose2015,Huang2015,Dhungana2016,Yuan2016}; these curves have been shifted in brightness to provide a reasonable match with the SN 2025pht curves. No correction for reddening has been applied to any of the curves.}
\label{fig:lc_opt}
\end{figure}

We show the BHTOM $BVI$ light curves for SN 2025pht in Figure~\ref{fig:lc_opt}. (These data are complete through 2025 October 26.) For comparison we also include light curves from several of the spectral comparison events: SN 2007od \citep{Inserra2011,Anderson2024}, SN 2008M \citep{Anderson2024}, SN 2013ej \citep{Valenti2014,Bose2015,Huang2015,Dhungana2016,Yuan2016}, and SN 2023ixf (\citealt{Zheng2025}; see also, e.g., \citealt{Teja2023,Hiramatsu2023,Singh2024,Li2025a}), as well as SN 2009bw \citep{Inserra2012}. Note that we made no further adjustments in time for the SN 2025pht curves, other than assuming our estimate of the explosion epoch, as we had inferred above from the spectral comparison. So, our choice of explosion epoch appears to be consistent between both the spectroscopic and photometric properties of the SN.

For SN 2025pht the photometry, then, commenced $\sim 20$ days after explosion, missing the peak, and followed a plateau in each band, much like the comparison SNe. The similarity in the post-peak inflection in the $B$-band curve is particularly striking. However, this is where the similarity with the evolution of most of the other SNe ends: SN~2025pht appeared to have fallen from the plateau after $\sim 78$--79 days, an interval shorter than the canonical $\sim 100$ days generally ascribed to SNe II-P, and potentially qualifying it as a ``short-plateau'' event \citep[e.g.,][]{Hiramatsu2021}. The post-plateau behavior of SN 2025pht appears to display a rather flat exponential tail, especially in $B$; the tail trend resembles fairly closely that of SN 2023ixf in $V$, but less so with $B$ and $I$. SN 2025pht photometrically most resembles SN 2008M at late times, although less so at earlier times along the plateau (we also only have data in $B$ and $V$).

If the peak brightness followed the same trends as the comparison SNe, SN 2025pht would have had a maximum brightness of $V\approx 13.4$ mag. Assuming the visual extinction and distance we have estimated below (in Sections~\ref{sec:reddening} and~\ref{sec:distance}, respectively), the peak absolute brightness would have been $M_V\approx -18.7$ mag, which is remarkably luminous.

\subsection{Photometry of the Progenitor Candidate}

The photometric analysis of the available {\sl HST\/} data followed the methods described by \citet{VanDyk2024}; in short, we pre-processed both the individual WFPC2 and WFC3 frames with {\tt Astrodrizzle} \citep[][with the additional benefit of flagging cosmic-ray hits in the data]{STScI2012b} and then measured the photometry with {\tt Dolphot} \citep{Dolphin2016}. The measurements were computed in Vega magnitudes and are listed as such in Table~\ref{tab:light_curve}; however, in Table~\ref{tab:observations} the brightness values were converted to AB magnitudes \citep{Oke1983}. Detection upper limits are at $5\sigma$ and were estimated in the same way and following the same logic as described by \citet{VanDyk2023}.

For the {\sl JWST \/} photometry we followed two different paths: we processed the data using the Python routine {\tt space{\textunderscore}phot}\footnote{https://space-phot.readthedocs.io/en/latest/} \citep{Pierel2024} and also with {\tt Dolphot} implemented for {\sl JWST}, following recommended parameter settings and methods from \citet{Weisz2024}. The resulting photometry is given in Table~\ref{tab:observations}. Note that the uncertainties in the {\tt Dolphot} measurements are underestimates, as \citet{Weisz2024} highlighted; we defer here from performing the necessary artificial star injection and recovery in each band that is essential for a more realistic uncertainty estimate. The agreement in the values from the two techniques is reasonably good, generally well within $2\sigma$ of the {\tt space{\textunderscore}phot} uncertainties, with the exceptions of the values at F277W and F770W ($\gtrsim 2\sigma$).
Given the uncertainty underestimates with {\tt Dolphot}, we adopted the {\tt space{\textunderscore}phot} measurements and more conservative uncertainties for the detections in our further analysis below (Section~\ref{sec:progenitor}).

The upper limit to detection in the F2100W band with {\tt space{\textunderscore}phot} was estimated via forced photometry at the SN position. Using {\tt Dolphot} we again estimated the upper limit at $5\sigma$ following the method we performed above for the {\sl HST\/} data.

We compare our photometric results with those of \citet{Kilpatrick2025}. The latter study used {\tt Dolphot} for photometry of both the {\sl HST\/} and {\sl JWST\/} data; thus, similar to our results, the uncertainties are quite likely underestimates. We find for the {\sl JWST\/} results that our {\tt Dolphot} values are on average $\sim 0.02$ mag brighter, and that the \citet{Kilpatrick2025} values are on average $\sim 0.05$ mag brighter than the {\tt space{\textunderscore}phot} values. The glaring exception, again, is at F770W, for which our measurements via {\tt Dolphot} and {\tt space{\textunderscore}phot} are 0.74 and 0.32 mag brighter than the \citet{Kilpatrick2025} measurements. In the case of F770W, as we have pointed out, the star sat on a ridge of likely PAH emission, so the underlying nebular background may have compromised clean photometry of the point source in general.

For the WFPC2 F814W measurements we find significantly larger differences: our {\tt Dolphot} photometry, after conversion to AB magnitudes, differ from the \citet{Kilpatrick2025} {\tt Dolphot} photometry by $\sim 3.2\sigma$ on average, relative to their uncertainties, and $\sim 8.6\sigma$, relative to our uncertainties, with our measurements being brighter (with the exception of the GO-9042 data for which the agreement is within the large uncertainties). The difference between our {\tt Dolphot} measurements from each of the two F814W frames per epoch differed at most by $2.4\sigma$, relative to the weighted uncertainty of the two, and were generally $\lesssim1\sigma$. The star is relatively isolated spatially, with regard to other sources in its vicinity, so it is difficult to understand these differences, based on the nature of the background. We speculate that the differences could arise from both the {\tt Astrodrizzle} pre-processing we undertook and in the assigned values of various {\tt Dolphot} input parameters\footnote{Among the more crucial {\tt Dolphot} parameters, for the WFPC2 photometry we assumed {\tt RAper}=8, {\tt FitSky}=3, {\tt img{\textunderscore}apsky}=15 25, {\tt RSky0}=15, {\tt RSky1}=35, {\tt ApCor}=1, {\tt WFPC2useCTE}=1, and {\tt InterpPSFlib}=1.}.

\subsection{Was The Progenitor Candidate Variable?}\label{sec:variability}

\begin{figure}
\plottwo{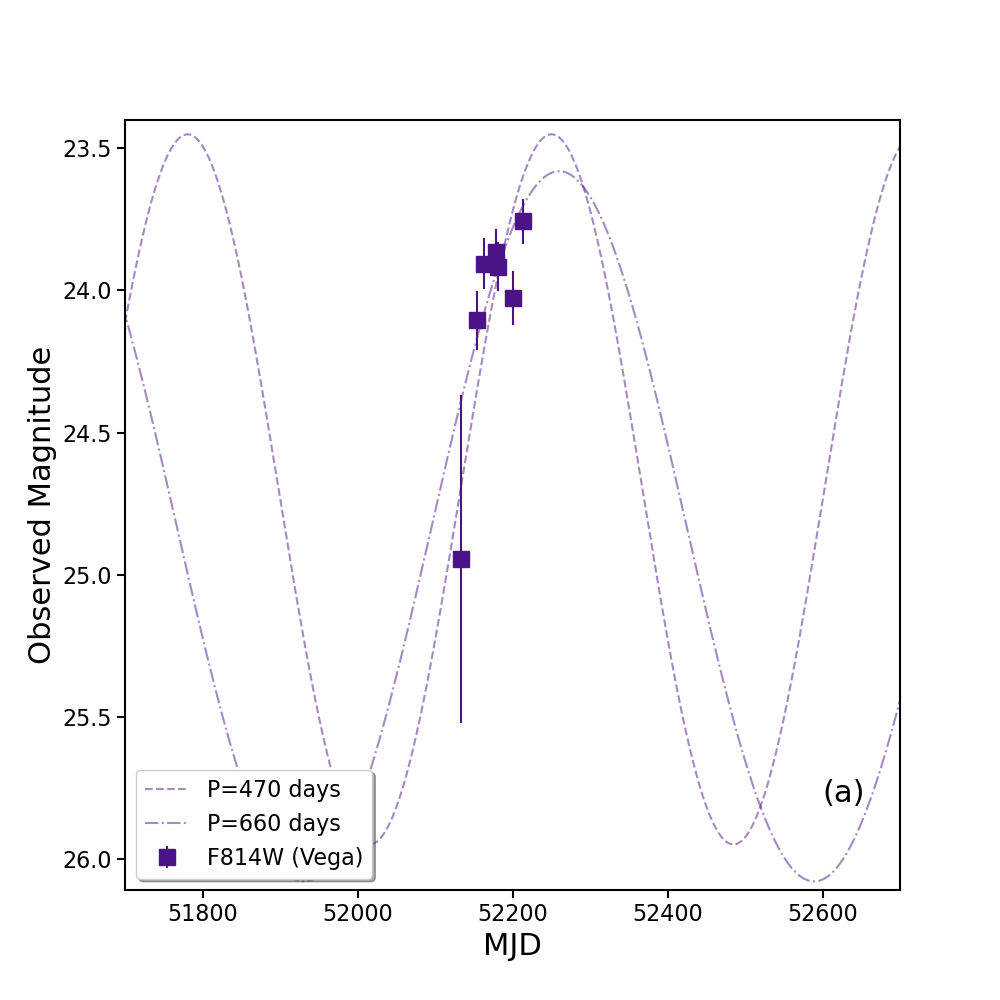}{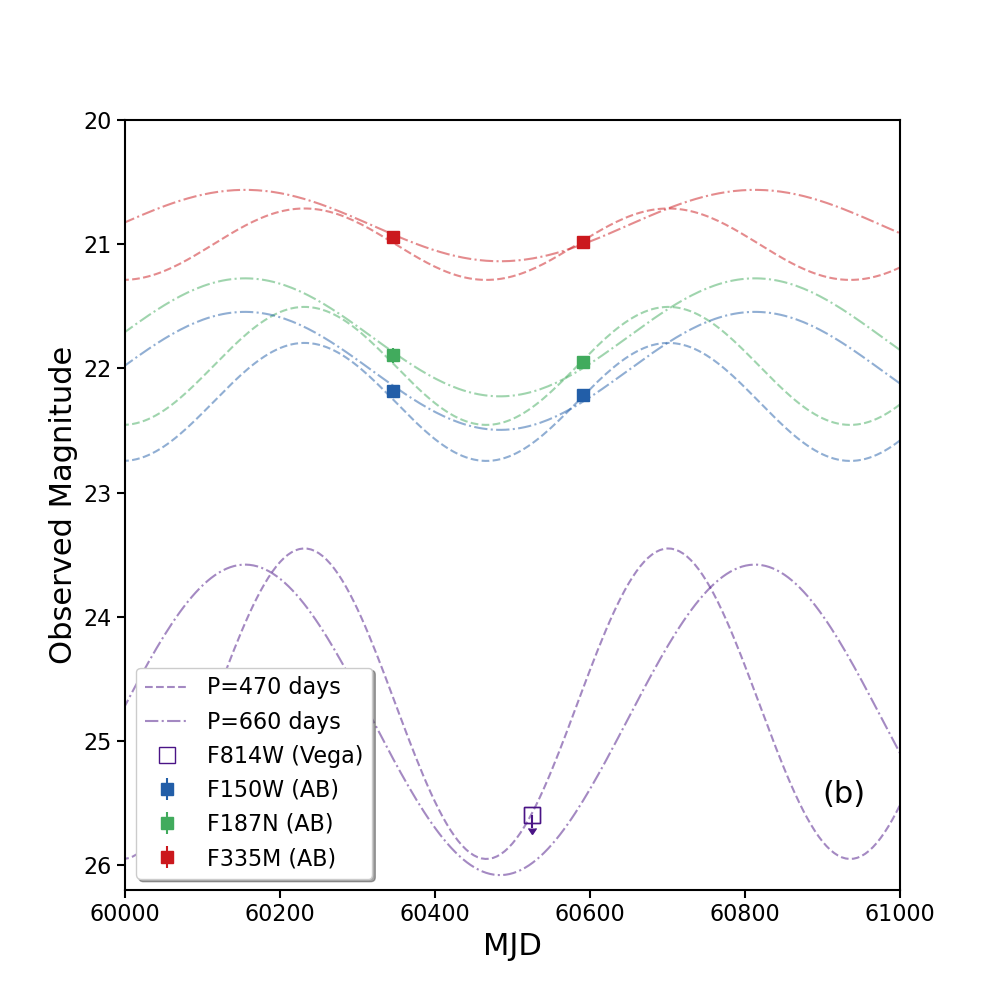}
\caption{(a): Photometry from {\sl HST\/} WFPC2 data from 2001 in F814W (see Table~\ref{tab:light_curve}). Also shown are simple sinusoid models with periods of 470 and 660 days. (b) Photometry from {\sl HST\/} WFC3/UVIS F814W and {\sl JWST\/} NIRCam F150W, F187N, and F335M data from 2024 (see Table~\ref{tab:observations}). Also displayed are the same sinusoid models shown in panel (a), which best fit the 2001 data, do not violate the 2024 F814W upper limit, and are in reasonable agreement with the NIRCam data.}
\label{fig:lc_f814w}
\end{figure}

The progenitor candidate was clearly detected in {\sl HST\/} F814W in 2001; see \citet[][their Figure~3]{Kilpatrick2025}. Examining Table~\ref{tab:light_curve}, it is evident that the star varied significantly in F814W brightness between 2001 August 12 and October 31 (the star was not detected at any time, either in 2001 or 2024, in F555W). In Figure~\ref{fig:lc_f814w}, panel (a), we show the F814W light curve during this time period. Visually one can see that the star was variable. However, clearly not enough data exist in this band (or at any other available wavelength, that we are aware of) to perform a rigorous period and amplitude analysis, similar to what, e.g., \citet{Jencson2023} and \citet{Soraisam2023} undertook for the SN 2023ixf progenitor candidate. However, we applied a very simple sinusoidal function with various values of the period and performed a reduced $\chi^2$ fit to the data points.

We further imposed the restriction that the model could not violate the upper limit on detection at F814W in 2024, and, to a lesser extent, that the model was in reasonable agreement with the brightnesses in the three NIRCam bands, F150W, F187N, and F335M, in which the host galaxy was observed by both {\sl JWST\/} programs. For the latter we considered the ratios of root-mean-square (RMS) amplitudes in the infrared, relative to the optical (\citealt{Soraisam2023}, for the SN 2023ixf progenitor candidate) and applied those to this star (we note that wavelength-dependent phase lags, by $\sim 0.1$--0.15, may also exist for variability in the infrared, relative to the optical, as \citealt{Smith2002} found for Mira variables). 

We found that this simple model met these requirements for sinusoids with periods of $\sim 470$ and of $\sim 660$ days, which both appear to provide a plausible representation of the F814W variability in 2001 and the NIRCam data; see Figure~\ref{fig:lc_f814w}. We were able to rule out models with periods $\gtrsim 1000$ days. The inferred peak amplitude at F814W would then be $\sim1.3$ mag, and therefore, the RMS amplitude is $\sim0.9$ mag, which is consistent with the star having been an LPV \citep[e.g.,][]{Soraisam2018}.

We therefore consider it compelling to conclude that the SN 2025pht progenitor candidate could have been a pulsational variable star. Admittedly, this is somewhat contrived and a bit fanciful, and we likely have already overinterpreted what minimal data we had at hand; note that, for this reason, we have not included any further complexity to the model, such as nonradial pulsations or higher-order overtones. Yet, the agreement between both the optical and {\sl JWST\/} light curves, such as they are, is at least self-consistent.

\subsection{Reddening To the Supernova}\label{sec:reddening}

The next step in the analysis is to estimate the total line of sight reddening to SN 2025pht, and then extrapolate this value to the total reddening to the progenitor candidate. We note first that the Galactic foreground visual extinction in the direction of the SN is $A_V=0.111$ mag ($E(B-V)=0.036$ mag; \citealt{Schlafly2011}, via NED). 

We have attempted to estimate the reddening via several different methods. (1) From the equivalent width (EQW)  of the Na~{\sc i}~D feature at the host-galaxy redshift in all of the SN 2025pht spectra (see Section~\ref{sec:naid} for details); (2) a comparison of the SN 2025pht spectra in their entirety with those of similar SNe II-P (see Section~\ref{sec:spectra} for details); and (3) via the $B-V$ color (see Section~\ref{sec:bminv} for details).

The Na~{\sc i}~D EQW has been empirically associated with the amount of reddening along a line of sight (\citealt{Munari1997,Turatto2003,Poznanski2012,Rodriguez2023}; however, also see \citealt{Phillips2013}). We estimated, based on \citet[][their Figure 9]{Phillips2013}, that the Na~{\sc i}~D EQW in the July 31 spectrum, when the feature is strongest, corresponds to $A_V=0.70\pm0.29$ mag. However, as those authors also emphasized and cautioned, the relation between EQW and $A_V$ becomes insensitive to dust extinction for EQW $\gtrsim 1$~\AA. Regardless, we consider that the presence of the Na~{\sc i}~D feature at all observed epochs qualitatively implies that a significant contribution to the SN reddening is in the host-galaxy reference frame. 

The comparison we then undertook was between the \citet{Strader2025} and Lick spectra (Figure~\ref{fig:spectracomp}) and those of SN~2007od \citep{Gutierrez2017}, SN 2008M (also \citealt{Gutierrez2017}), and SN 2013ej \citep{Dhungana2016}, which either were not host-reddened or could be corrected for host reddening reasonably well. From the SN~2025pht spectra for all four dates, taken at face value, the average amount of additional (``addl'') reddening required for the comparison spectra to fit the spectral shape for SN 2025pht, $E(B-V)_{\rm addl}$, is $0.55\pm0.05$ mag.

Lastly, we considered the $B-V$ color of SN 2025pht, and compared it to that of SN 2007od \citep{Inserra2011,Anderson2024}, SN 2008M \citep{Anderson2024}, SN 2009bw \citep{Inserra2012}, SN 2013ej \citep{Valenti2014,Bose2015,Huang2015,Dhungana2016,Yuan2016}, and SN~2023ixf \citep{Zheng2025}. Although \citet{deJaeger2018} argued that significant dispersion exists across SN~II intrinsic color, independent of any host-galaxy reddening, and therefore no true ``color template'' can possibly be constructed, we have fit the colors of these comparison SNe to that of SN 2025pht. To fit the SN 2025pht ($B-V$) color the average for the five comparison SNe is $E(B-V)_{\rm addl}=0.54 \pm 0.12$ mag, quite consistent with the estimate from the spectral comparison.

None of these methods is particularly foolproof. Ultimately, though, we considered the average color excess across both the spectral and color-curve comparisons and assume hereafter that $E(B-V)_{\rm host}=E(B-V)_{\rm addl}=0.54\pm0.08$ mag. Assuming $R_V=3.1$ typical for interstellar dust, the visual extinction $A_V({\rm host})=1.69\pm0.25$ mag. (We note that the value of $A_V[{\rm host}]$ appears to be clearly underestimated by the Na~{\sc i}~D EQW technique.) Together with the Galactic $A_V$, this is $A_V({\rm tot})=1.80\pm0.25$ mag. The SN has evidently experienced a considerable amount of interstellar dust extinction along our line of sight.

We hereafter assume that this total extinction to the SN is applicable to the progenitor candidate as well. With regard to the candidate photometry, the {\sl HST\/} bands are, of course, affected the most by this high total extinction, from $A_{\lambda} \approx 1.09$ to 3.63 mag. The {\sl JWST\/} bands are far less affected, from $A_{\lambda} \approx 0.07$ to $\sim 0.38$ mag, with the least extinction in MIRI F770W ($A_{\lambda} \approx 0.03$). For the optical through NIRCam bands we have assumed the \citet{Cardelli1989} reddening law. For the F2100W band we assumed the reddening law from \citet{Xue2016}.

\subsection{Distance to the Host Galaxy}\label{sec:distance}

A number of distance estimates exist for the host galaxy, NGC 1637. Most of these are based on measurements associated with SN 1999em. \citet{Hamuy2001}, \citet{Leonard2002}, and \citet{Elmhamdi2003} all established estimates using the expanding photosphere method (EPM) and arrived at quite similar values: $7.5 \pm 0.5$, $8.2 \pm 0.6$, and $7.83 \pm 0.3$ Mpc, respectively.
\citet{Baron2004} employed a spectral-fitting ``expanding atmosphere method'' and found a substantially larger $12.5\pm 1.8$ Mpc.
\citet{Jones2009} recomputed the EPM distance assuming two different sets of SN~II atmosphere models: $9.3\pm0.5$ Mpc from the \citet[][ E96]{Eastman1996} models and (a quite different) $13.9\pm1.4$ Mpc using the \citet[][ D05]{Dessart2005} models.
Via quantitative spectroscopic analysis, \citet{Dessart2006} arrived at a distance of $11.5\pm1.0$ Mpc.
\citet{Olivares2008} applied the standard candle method (SCM) to SN 1999em and found $10.0\pm1.3$ Mpc. 

\citet{Leonard2003} measured a Cepheid-based distance of $11.7\pm 1.0$ Mpc from the {\sl HST\/} WFPC2 observations of NGC 1637 in 2001 listed in Table~\ref{tab:light_curve}. \citet{Saha2006} measured, from a similar technique employing the same data, $d=12.02\pm0.39$ Mpc, which agrees with the \citet{Leonard2003} measurement to within the uncertainties.

We estimated another SCM distance to SN 1999em adopting the calibration by \citet{Polshaw2015}, which is anchored to the megamaser galaxy NGC 4258 (M106); see Section~\ref{sec:SCM_99em}. Our result is $d=10.45\pm2.32$ Mpc, based on the expansion velocity from the Fe~{\sc ii} $\lambda$5169 absorption line feature, and $d=9.86\pm1.82$ Mpc, from the weighted average of the velocities from Fe~{\sc ii} $\lambda$5169, $\lambda$4629, $\lambda$4924, and $\lambda$5018 lines. These distance estimates agree with each other to within the (large) uncertainties. They also agree with the SCM distance by \citet{Olivares2008}, again, to within the uncertainties. At the very least, we have provided a cross-check on the previous estimate, employing a different SCM calibration.
We also measured a distance estimate to the host galaxy from SCM for SN 2025pht, $d=8.65\pm0.88$ Mpc. See Section~\ref{sec:SCM_25pht}. This SCM distance is on the comparatively low side, although it agrees to within the uncertainties with the SCM distances from SN 1999em both in this study and by \citet{Olivares2008}. It also agrees with the early EPM distances to SN 1999em as well.

Finally, we exploited the fact that we have {\sl JWST\/} photometry not only for the progenitor candidate, but for the entire host galaxy that was encompassed on the NIRCam detectors. This is especially relevant for measuring distances employing a relatively new technique, based on the presence of the J-region asymptotic giant branch (AGB), dominated by carbon-rich AGB stars and best detectable in the near-infrared \citep[e.g.,][]{Madore2020,Freedman2020,Lee2025,Li2024,Li2025b}, which agrees well with other distance indicators, such as the tip-of-the-red-giant branch (TRGB) method (\citealt{Freedman2025}; although see \citealt{Anand2025}); details are given in Section~\ref{sec:jagb}. From this technique we found a distance $d=11.38\pm0.58$ Mpc.

\begin{figure}
\plotone{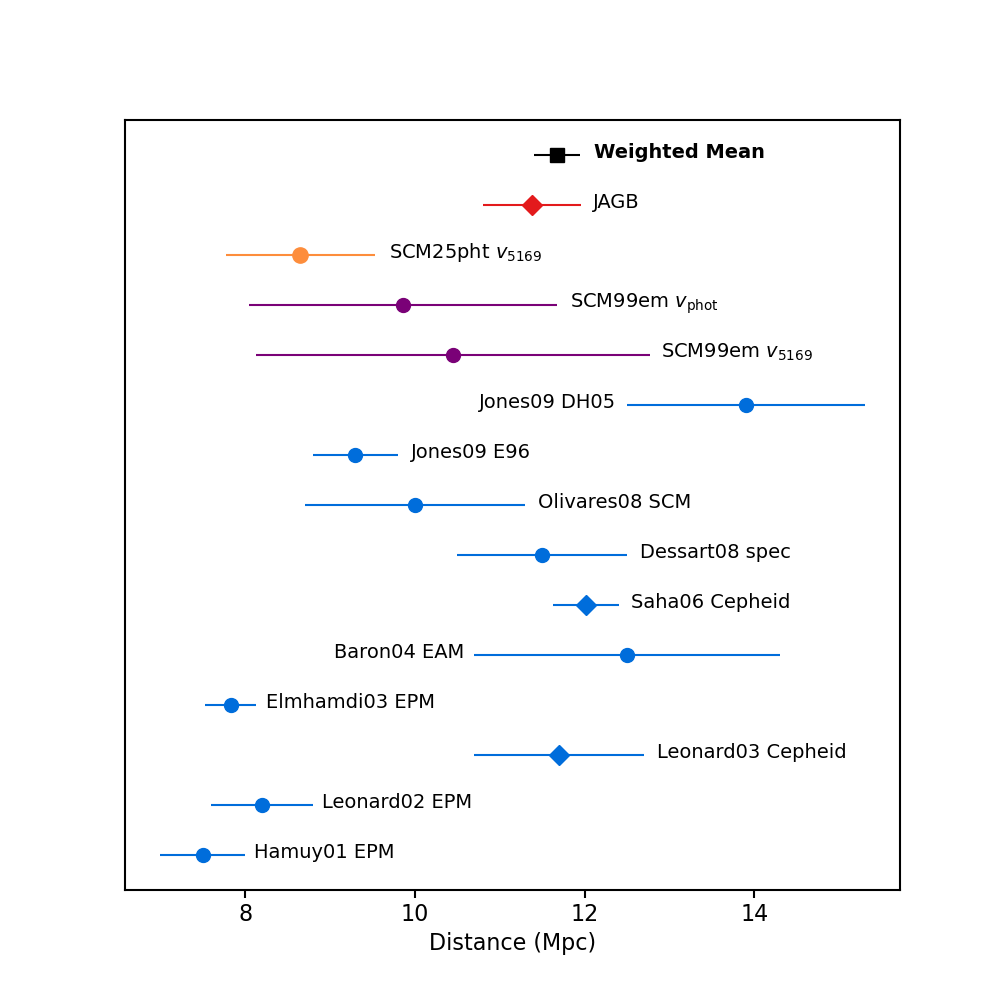}
\caption{A compendium of all of the known distance estimates to the host galaxy NGC 1637. See text. The SN-based estimates we have culled from the literature are shown as blue filled circles. Also included here are estimates we have made in this study, based on SCM measurements to SN 1999em (purple filled circles) and SN 2025pht (orange filled circle). Cepheid-based measurements from the literature are shown as blue filled diamonds. A JAGB-based measurement from the archival {\sl JWST\/} observations of the host that we have made here is shown as a red filled diamond. In addition, we show the weighted mean of the galaxy-based distances we have considered here (filled square).
\label{fig:distances}}
\end{figure}

In summary, we consider all of these various distance estimates taken together, as shown in Figure~\ref{fig:distances}. They span a large range, as can be seen. The three early SN 1999em-based EPM estimates are all on the low end of the range. Further later considerations of SN 1999em as a distance indicator resulted in somewhat larger distances, including the SCM estimates both by \citet{Olivares2008} and this work. Our attempt here at an SCM distance for SN 2025pht is statistically consistent with those for SN 1999em; however, taken on its own, it is more indicative of a shorter distance. We had originally computed an uncertainty-weighted mean of all of the known distances to the host galaxy. However, the SN-based techniques are likely to be less certain, since individual SNe II-P can vary one from the other, and the calibrations of these distance indicators are less well established. Whereas, the galaxy-based indicators, which tend to indicate a longer distance, are likely more reliable. We subsequently therefore established a weighted mean from the latter indicators only, $11.67\pm0.27$ Mpc ($\mu=30.34\pm0.05$ mag), which we adopt hereafter. It is interesting that the \citet{Leonard2003} Cepheid-based measurement and that from the JAGB conducted here agree with each other well to within their respective uncertainties. Note that \citet{Kilpatrick2025} assumed the \citet{Saha2006} distance, which agrees with our adopted value to within the uncertainties.

As a final note here, it should be possible in principle to consider the extensive WFPC2 data from programs GO-9042 and GO-9155 in F555W and F814W, i.e., in $\sim I$ and $\sim (V-I)$, to isolate the TRGB in these data \citep[e.g.,][]{Lee1993,Freedman2019,Anand2021}.
Additionally, we could  consider the WFC3 data from GO-17502 in these same bands, to take advantage of the superior sensitivity and resolution.
Unfortunately, as we show in Section~\ref{sec:trgb}, the resulting color-magnitude diagram (CMD) from both datasets is not deep enough and is too confused to reveal the TRGB.
Furthermore, \citet{Newman2024} provided calibrations in {\sl JWST\/} NIRCam bands for the TRGB; however, again unfortunately, none of the combinations of bands presented there are represented in the available archival data here.

\section{The Inferred Properties of the Progenitor Candidate}\label{sec:progenitor}

\begin{figure*}
\plottwo{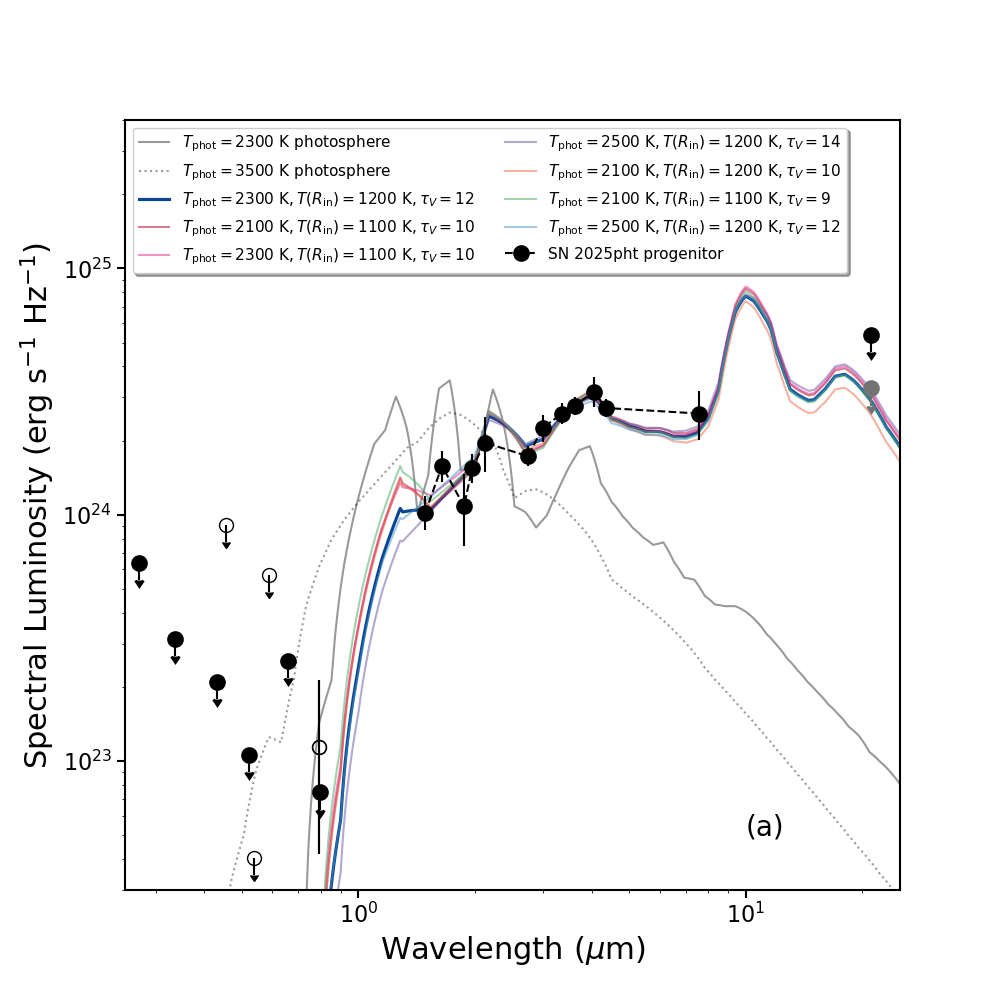}{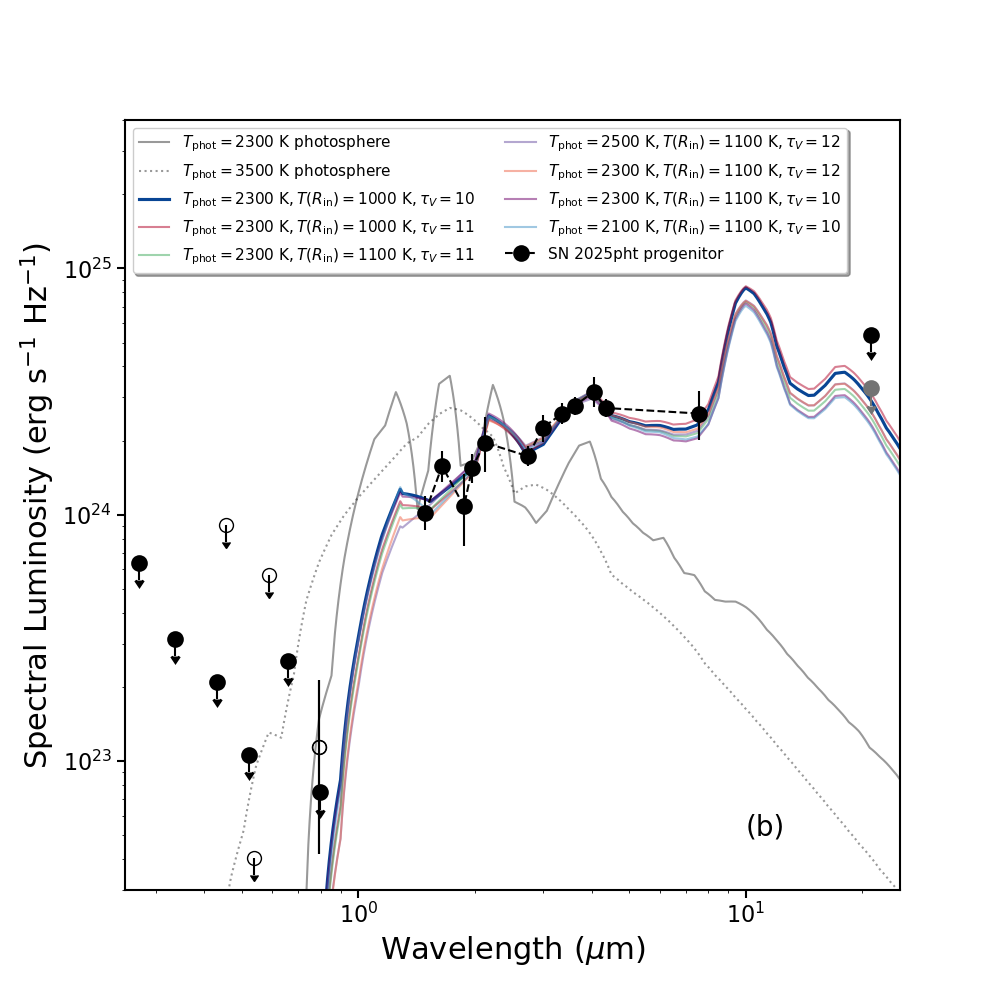}
%\gridline{\fig{sn25pht_fitting_density1.png}{0.50\textwidth}{}}
%\gridline{\fig{sn25pht_fitting_density4.png}{0.50\textwidth}{}}
\caption{The reddening- and distance-corrected observed SED of the SN 2025pht progenitor candidate. Measurements from {\sl HST\/} and {\sl JWST\/} data from 2024 are shown with solid circles; the {\sl HST\/} data from 2001 are open circles. Only the faintest {\sl HST\/} F814W detection from 2001 (see Table~\ref{tab:light_curve} and Figure~\ref{fig:lc_f814w}) is shown. The upper limit to detection in {\sl JWST\/} MIRI F2100W estimated from {\tt Dolphot} \citep{Dolphin2016} processing is indicated with a black circle, whereas that using {\tt space{\textunderscore}phot} is a dark gray circle. Shown for comparison are various model SEDs generated using {\tt DUSTY} \citep{Ivezic1997,Ivezic1999,Elitzur2001} --- with a range of photospheric temperature, $T_{\rm phot}$, temperature at the inner radius of the spherical model dust shell, $T(R_{\rm in})$, and dust optical depth at 0.55 $\mu$m, $\tau_V$ --- which fit well the observed SED. Also shown is the SED of the {\tt PHOENIX} model photosphere \citep{Kucinskas2005,Kucinskas2006} at 2300 K input to {\tt DUSTY}, as well as a {\tt PHOENIX} atmosphere at 3500 K, scaled to the same bolometric luminosity as the best-fit {\tt DUSTY} model. (a) Models generated assuming a single shell with {\tt density type=1}; and (b) models generated with {\tt density type=4}. See text.
\label{fig:prog_fit}}
\end{figure*}

In Figure~\ref{fig:prog_fit} we show the reddening- and distance-corrected spectral energy distribution (SED) in all of the available {\sl HST\/} and {\sl JWST\/} bands for the SN 2025pht progenitor candidate. We are fortunate in that all of the NIRCam and MIRI observations, as well as all of the WFC3 observations, are within $\sim$246 days ($\sim$0.67 yr) of each other in calendar 2024. So, we have what constitutes a ``quasi-snapshot'' of the spectral luminosity of the star, unlike the majority of other previous progenitor identification attempts.
The three bands in common between the two {\sl JWST\/} datasets (F150W, F187N, and F335M) differ in brightness by $<0.05$ mag between the two observation dates (Table~\ref{tab:observations} and Figure~\ref{fig:lc_f814w}). Thus, we can assume that it is reasonable to combine all of the {\sl JWST\/} bands together.

We also show in Figure~\ref{fig:prog_fit} a {\tt PHOENIX} model photospheric SED at solar metallicity with surface gravity $\log (g\ [{\rm cm\ s}^{-2}])=-0.5$ \citep{Kucinskas2005,Kucinskas2006}, at a temperature of 2300 K. This model obviously does not represent the observed SED at all well. (One would have to opt for a much cooler photosphere or blackbody to even approximate the SED, let alone fit it.) So, clearly what we are seeing in the case of the SN progenitor candidate is emission from a dusty CSM, which was reprocessing the (primarily ultraviolet through near-infrared) light of the central stellar photosphere via dust radiative transfer into the mid-infrared and, assumedly, even longer wavelengths.

For that reason, we have generated models for the SED using {\tt DUSTY} \citep{Ivezic1997,Ivezic1999,Elitzur2001}. For the properties of the simple spherical circumstellar dust-shell model we have assumed generally that the composition is O-rich ``warm'' silicates \citep{Ossenkopf1992}. \citet{VanDyk2024} argued that, although the observed SEDs of some Galactic RSGs can be reproduced with the addition of a small mass fraction ($\lesssim$5\%) of amorphous carbon dust mixed in with the predominantly Si dust mass \citep{Verhoelst2009}, an assumption of pure, or near-pure, C-rich dust in this modeling is unjustified. (Any suggestion of carbon grains, such as PAH emission, would likely be ambient interstellar matter in front of, swept up by, or otherwise affected, e.g., irradiated, by the star.) We have also assumed a ``modified Mathis--Rumpl--Nordsieck'' \citep[MRN;][]{Mathis1977} dust-grain size distribution (i.e., index $q=3.5$ and grain size range $a({\rm min})=0.005$ to $a({\rm min})=0.25\ \mu$m). The  photospheric SEDs input into {\tt DUSTY} at the center of the model shell are the {\tt PHOENIX} models, mentioned above. 

We had originally assumed a dust density distribution for the models to be a single shell with {\tt DUSTY} parameter {\tt density type=1 (N=1)}. The spherical shell has an inner radius $R_{\rm in}$, at which the dust grains have condensed, and the dust density decreases $\propto r^{-2}$, i.e., a steady-state wind (mass-loss rate $\dot M=$ const.) with constant velocity, to an outer radius $R_{\rm out}=10^3\ R_{\rm in}$. We subsequently also assumed the analytical approximation for radiatively-driven winds, with $R_{\rm out}=10^4\ R_{\rm in}$, i.e., {\tt DUSTY} parameter {\tt density type=4}. Although the wind driving mechanism is considered uncertain for RSGs \citep{Beasor2022}, both assumptions have been adopted by various investigators modeling RSG dust emission (steady-state, e.g., \citealt{Fok2012,Humphreys2020}; radiatively-driven, e.g., \citealt{vanLoon2005,Antoniadis2024}).

The models are then determined solely by the stellar photospheric temperature ($T_{\rm phot}$), the temperature at the inner radius of the dust sphere ($T[R_{\rm in}]$), and the dust optical depth ($\tau_V$) at 0.55 $\mu$m. We applied a reduced chi-squared fitting, $\chi_{\rm red}^2\le 1$, of a relatively coarse grid of models to the star's SED. We constrained the model fitting at long wavelengths based on the upper limit at F2100W, primarily the limit from {\tt space{\textunderscore}phot}, which is stricter than that from {\tt Dolphot}. The SED is essentially unconstrained at short wavelengths ($< 1.5\ \mu{\rm m}$). For {\tt density type=1} we found the allowed ranges in the input parameters to be limited to $T_{\rm phot}=2100$--2500 K, $T(R_{\rm in})=1100$--1200 K, and $\tau_V=9$--14. For {\tt density type=4} we found a similar range of $T_{\rm phot}=2100$--2500~K, $T(R_{\rm in})=1100$--1200 K, and $\tau_V=10$--12. (The preeminent difference between the two sets of models is a narrower allowed range in $\tau_V$ for {\tt density type=4}.) We show in Figure~\ref{fig:prog_fit} a number of the allowed models for both density distributions. The two sets of model SEDs have quite similar overall shapes. Although these models do not precisely follow the shape of the observed SED at (for example) F164N and F187N, it is our view that overall they provide quite satisfactory representations of the data to within the uncertainties. 

Interestingly, the {\sl HST\/} data provide no meaningful constraints on the nature of the observed SED, particularly at F814W. In this case of SN 2025pht, the progenitor candidate SED is, for the first time, entirely defined by {\sl JWST\/} observations obtained fortuitously not long before explosion. Also for the first time, the SED extends to the red beyond ({\sl Spitzer}) 4.5 $\mu$m, which was the limit for SN 2023ixf.

We also note that we could add no more than $\sim4$\% of amorphous C \citep{Preibisch1993} and still fit the observed SED to within the uncertainties. The predominantly Si-rich nature of the CSM dust appears to be quite firm.

The best-fit model SED for {\tt density type=1} had $T_{\rm phot}=2300$~K, $T(R_{\rm in})=1200$~K, and $\tau_V=12$. Integrating over that model (from 0.25 to 121,000 $\mu$m) the resulting bolometric luminosity is $L_{\rm bol}=5.22\ ({\pm 0.32})\times10^{38}$ erg s$^{-1}$, or $\log(L_{\rm bol}/L_{\odot})=5.13 \pm 0.03$. The best-fit model SED for {\tt density type=4} had $T_{\rm phot}=2300$~K, $T(R_{\rm in})=1000$~K, and $\tau_V=10$, and integrating over that model resulted in a somewhat higher $L_{\rm bol}=5.49^{+0.22}_{-0.66}\times10^{38}$ erg~s$^{-1}$, $\log(L_{\rm bol}/L_{\odot})=5.16 \pm 0.03$, although the two luminosity values agree to within their uncertainties. We note that if the dust geometry were asymmetrical or non-spherical, these luminosities may be overestimates, as is likely the case for the extreme RSG WOH G64 in the Large Magellanic Cloud (LMC), for which the circumstellar dust appears to be in a torus \citep{Ohnaka2008}.

The effective radius of the star, $R_{\rm eff}$, is $\sim 2327 \pm 70\, R_{\odot}$ and $\sim 2385 \pm 72\, R_{\odot}$ for the best-fit {\tt density type=1} and {\tt density type=4} models, respectively, which would be truly enormous; however, the large computed $R_{\rm eff}$ is primarily a result of the assumed low $T_{\rm eff}$. As far as the dust shell is concerned, at these $L_{\rm bol}$, $R_{\rm in} \approx (5.76 \pm 0.17) \times 10^{14}$ cm, or $\sim 3.6\, R_{\rm eff}$, and $R_{\rm in} \approx (9.34 \pm 0.29) \times 10^{14}$ cm, or $\sim 5.7\, R_{\rm eff}$, respectively --- although as the {\tt DUSTY\/} output warns, in compliance with the point-source approximation this result is only really applicable for $T_{\rm eff}>2508$ K, and should therefore be taken with a grain of salt (... or dust). The $R_{\rm in}$ values that we arrived at here are smaller than what is typically expected for RSGs ($R_{\rm in} \gtrsim 10\ R_{\rm eff}$, e.g., \citealt{Verhoelst2009}; although, the inner dust radius for the less-enshrouded Betelgeuse may only be $\sim 1.5$--$2.5 \times$ the stellar radius, \citealt{Haubois2019}). A factor that could also be leading to an overestimate here of $R_{\rm eff}$ is our simple model assumption of spherical symmetry for the dust shell, whereas circumstellar environments can be asymmetric or clumpy, or both \citep[e.g.,][]{Smith2001,Ohnaka2008,Shenoy2016,Ma2025}.

From the best-fit models the visual extinction along the line of sight from CSM dust would be $A_V({\rm CSM})\approx7.6$ and $\approx 6.3$ mag, respectively (for comparison, we estimated in Section~\ref{sec:reddening} that $A_V[{\rm host}]=1.69$ mag).

The range of apparently very cool input photospheres could be a result of the high dust obscuration by the CSM ---  \citet{VanDyk2024} pointed to the example of the extremely dusty star VY Canis Majoris, for which interferometric observations indicate a far warmer photosphere \citep[$\sim 3490$ K,][]{Wittkowski2012} than does SED modeling ($\sim 2800$ K, \citealt{Harwit2001}; although see \citealt{Massey2006}). We acknowledge that it is very difficult to determine accurately a $T_{\rm eff}$ from the SED alone in the presence of circumstellar dust. However, the best fitting models tend to follow quite well the shapes of both the observed SED between 1.5 and 4.5 $\mu$m, in particular, and the molecular absorption features in the low-temperature (e.g., 2300~K) photosphere. (These observed and model undulations in the SED in that wavelength range are not inherent to the O-rich, Si dust component; see \citealt{Ossenkopf1992}.) It is also difficult to envision how the reprocessed light from, say, a warmer 3500 K photosphere (see Figure~\ref{fig:prog_fit}) could behave in this same manner.

We note that, should the photospheric temperature actually be closer to 3500~K, then $R_{\rm eff} \approx 941\ R_{\odot}$, which is thought to be more typical for RSGs, based on previous estimates for these stars across nominally a broad luminosity range. Furthermore, for the dust shell, $R_{\rm in} \approx10.1\ R_{\rm eff}$ (and the temperature caveat from {\tt DUSTY} would not apply in this case).

\begin{figure}
\plotone{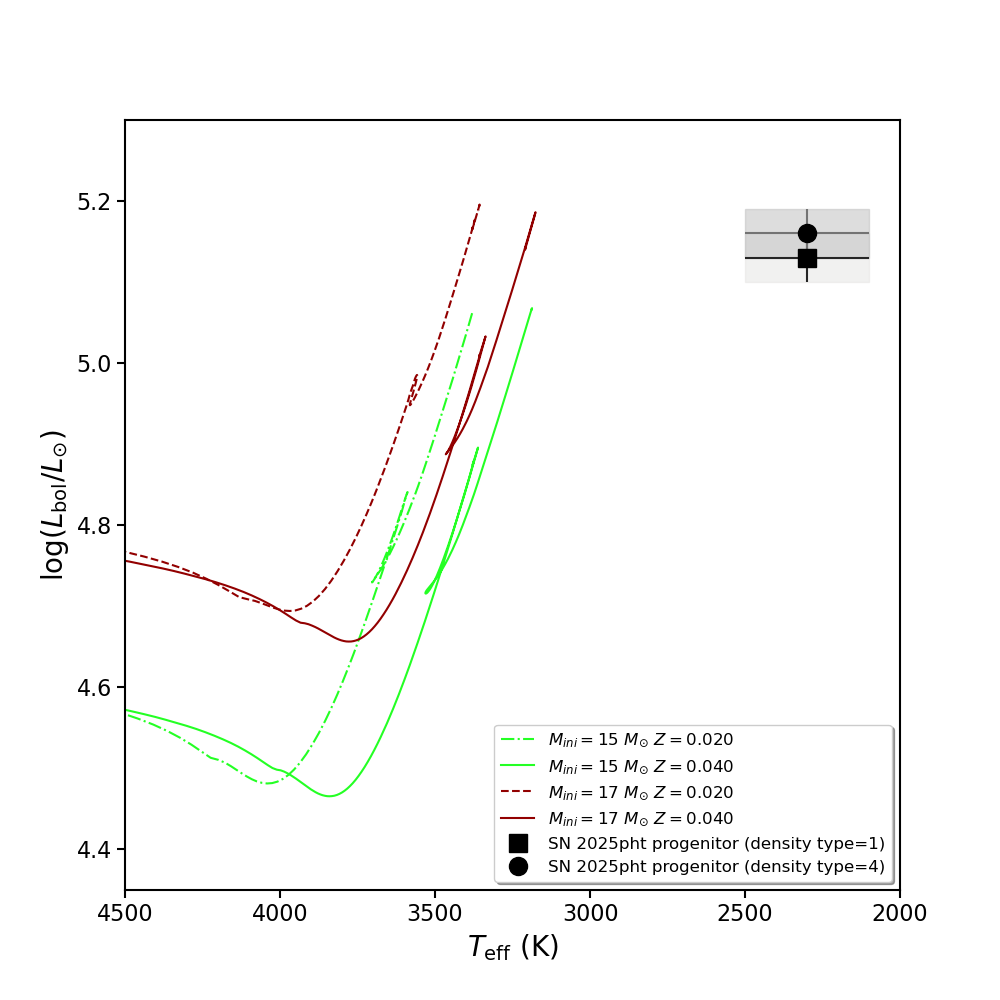}
\caption{Hertzsprung-Russell diagram showing the inferred locus in $T_{\rm eff}$ and $L_{\rm bol}$ of the SN 2025pht progenitor candidate (black square). Also shown are theoretical BPASS single-star evolutionary tracks \citep{Stanway2018} at two different metallicities, $Z=0.020$ (approximately solar) and $0.040$ (above solar), at $M_{\rm ini}=15$ and $17\ M_{\odot}$.
\label{fig:hrd}}
\end{figure}

As \citet{VanDyk2025} highlighted, connecting RSG $L_{\rm bol}$ to $M_{\rm ini}$ using theoretical evolutionary tracks introduces further uncertainties and is open to interpretation \citep[see][]{Davies2020} --- and additionally, \citet{Laplace2026} have advised against inferring $M_{\rm ini}$ from a single, pre-SN estimate of $L_{\rm bol}$ and $T_{\rm eff}$ for RSGs that may have experienced high-amplitude pulsational variability prior to explosion\footnote{This realization is sobering, since one of the primary aims of progenitor identification, up to this point at least, has been defining the range in $M_{\rm ini}$.} --- nevertheless, we show in Figure~\ref{fig:hrd} a Hertzsprung-Russell diagram with the locus of the progenitor candidate, based on the best-fit SED model. We also show for comparison BPASS single-star evolutionary tracks \citep{Stanway2018} at two different metallicities, $Z$. 

As can be seen in the figure, none of the tracks shown terminate at a cool enough temperature to match the inferred $T_{\rm eff}$ of the progenitor candidate. (We have discussed the potential temperature disparity, above.)
RSGs are known with cooler photospheres than model evolutionary tracks can reproduce, e.g., WOH G64 \citep{Levesque2009}. This results from a combination of mantle depletion and expansion, due to mass loss, and the pulsationally-induced increase of the atmospheric scale height; the luminosity of the star, and the model, are determined by the core mass and the nuclear physics occurring there \citep{vanLoon2025}. The range in estimated $L_{\rm bol}$, at least from this apparition of the candidate in 2024, therefore is consistent with the tracks generated between approximately $M_{\rm ini}=15$ and $17\ M_{\odot}$. This mass range is consistent with what has been inferred for other SN II-P progenitors, albeit somewhat on the higher side. In fact, the high end of this range pushes up against the original observational limit placed on the highest-mass progenitors --- the origin of the ``RSG problem'' \citep{Smartt2009}.

\section{Discussion and Conclusions}\label{sec:discussion}

A full analysis of SN 2025pht itself is beyond the scope of this work; however, we can say that we find it likely that it is a short-plateau Type II-P event. Such SNe are also quite luminous, surrounded by dense CSM resulting from high mass loss, and may generally arise from high-mass progenitor stars (\citealt{Hiramatsu2021}; although see \citealt{Teja2022}). We would likely expect these SNe, given the massive CSM set up prior to explosion, to exhibit sustained emission at late times as a result of SN shock-CSM interaction.

We have also inferred that the star that gave rise to SN 2025pht was, {\em at least in 2024}, a luminous RSG surrounded by dusty CSM. The star is also the candidate, to our knowledge, most extinguished by interstellar dust within the host galaxy. Without the archival {\sl JWST\/} observations we would not have been able to fully characterize (or even potentially detect) the candidate in the first place. The only information that we would have had, based solely on the {\sl HST\/} F814W data, was that the star was variable in brightness, possibly as an LPV. However, that alone would not have necessarily provided any further indications regarding the properties of the star. 

\begin{figure}
\plotone{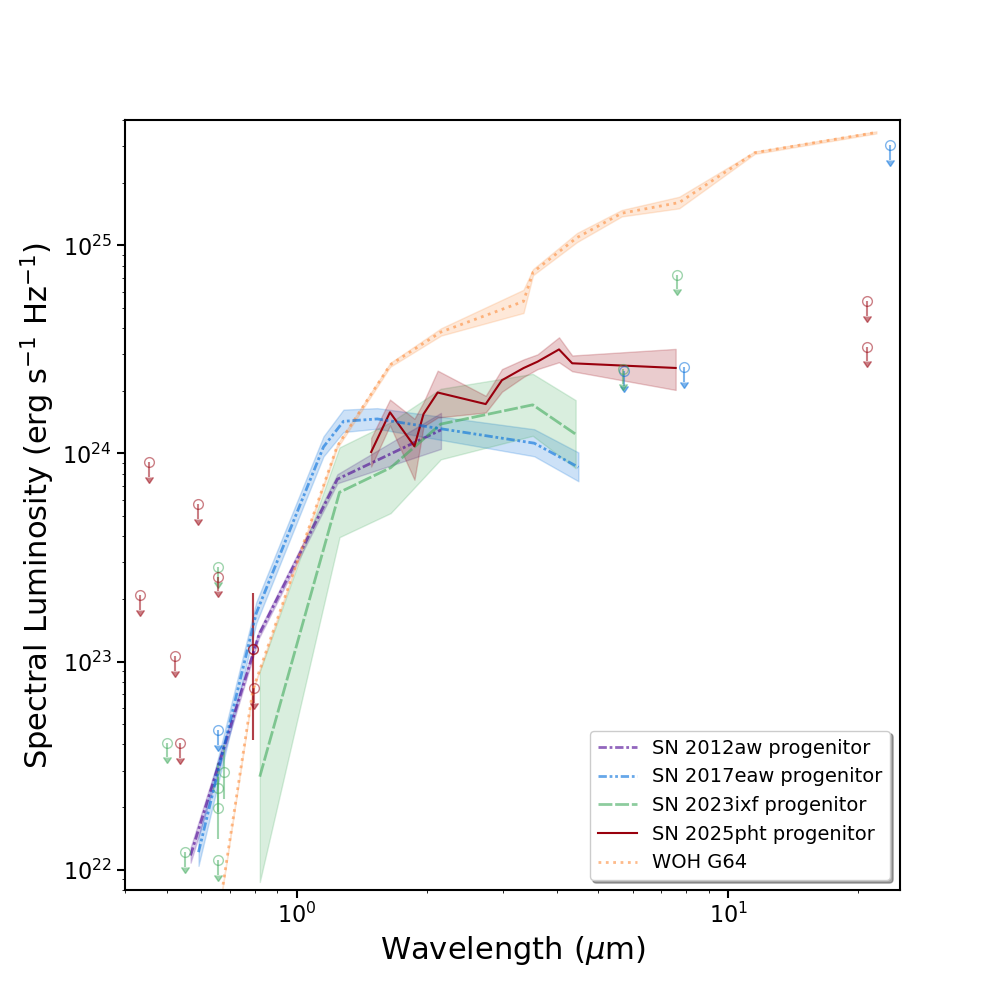}
\caption{Comparison of the observed, reddening- and distance-corrected, SEDs of four of the dustiest known SN progenitor candidates: SN 2025pht (this study), SN 2012aw \citep{VanDyk2012}, SN 2017eaw \citep{VanDyk2019}, and SN 2023ixf \citep{VanDyk2024}. Both the candidates for SN 2012aw \citep{VanDyk2015,Fraser2016} and SN 2017eaw \citep{VanDyk2023} have since been confirmed; the candidate confirmations for both SN 2023ixf and SN~2025pht remain pending. (We also show for comparison the SED of the extreme RSG WOH G64;  \citealt{Ohnaka2008,Levesque2009,Ohnaka2024,vanLoon2026}.)
\label{fig:sn_prog_comp}}
\end{figure}

The star is among the most luminous (if not {\em the\/} most luminous) RSG, at a luminosity possibly as high as $\log(L_{\rm bol}/L_{\odot})=5.16 \pm 0.03$, that has been identified so far as an SN II-P progenitor candidate. %At the upper boundary of this estimate, this pushes 
This begins to push somewhat against the ceiling set by the RSG problem. It also may well be the dustiest, or certainly among the dustiest, progenitors. See Figure~\ref{fig:sn_prog_comp}, in which we compare the SN 2025pht progenitor candidate observed SED with the dusty, luminous progenitor SEDs of other SNe II-P, specifically SN 2012aw (\citealt{VanDyk2012}; also \citealt{Kochanek2012,Fraser2012}), SN 2017eaw (\citealt{VanDyk2019}; also \citealt{Kilpatrick2018,Rui2019}), and SN 2023ixf (\citealt{VanDyk2024}; also, e.g., \citealt{Kilpatrick2023,Jencson2023,Xiang2024,Ransome2024,Qin2024}). We also provide for comparison the SED of the extreme RSG WOH G64 (IRAS 04553$-$6825, \citealt{Levesque2009,Ohnaka2024,vanLoon2026}; $\log(L_{\rm bol}/L_{\odot})=5.45$, \citealt{Ohnaka2008}). The SN 2025pht candidate SED differs notably from the SN 2017eaw one; however, it is similar in shape and overall luminosity with that for SN 2023ixf (despite the apparent lack of similarity in the spectra and light curves of SN 2025pht and that SN), although the latter does not extend past 4.5 $\mu$m. We do not really know the overall shape of the SN 2012aw progenitor candidate SED, since it was only definable to about 2 $\mu$m. That our measured SED for the SN 2025pht candidate is overall similar to these other SN progenitors, particularly to that of SN 2023ixf, provides us with some satisfaction in the veracity of our analysis here.

From {\tt DUSTY} modeling of the SN 2023ixf progenitor candidate, following similar input assumptions, \citet{VanDyk2024} arrived at best-fitting models with ranges $T_{\rm eff}=2300$--2700 K, $T(R_{\rm in})=900$--1300 K, and $\tau_V=9$--11, which are quite similar to what we found here for the SN 2025pht candidate. %(Could the inferred cooler-than-expected $T_{\rm eff}$ in both cases be an evolutionary state prior to explosion?) 
Yet, the two SNe, as previously mentioned, appear to have noteworthy differences in their properties and evolution, even if the two RSG progenitors may have had similarities in their nature, possibly pointing to overall differences in mass-loss histories, progenitor environments, and interactions with a possible binary companion. The dependence of the nature of observed SNe on pre-explosion mass loss cannot be overstated \citep[e.g.,][]{vanLoon2025}. Since SN 2025pht was discovered $\sim 14$ days post-explosion, and the first spectrum was obtained at $\sim 18$ days, any flash-ionization features that may have appeared at early times likely were no longer detectable. So, there is no way of determining the nature of the CSM immediate to the progenitor and early shock interaction with it, and to compare that with (for example) the early evolution of SN 2023ixf (for which the flash features persisted for $\sim 7$ days, implying a shock radius of $\sim 6 \times 10^{14}$~cm; \citealt{JacobsonGalan2025}, and references therein).

Although we are in agreement with \citet{Kilpatrick2025} that the star was likely highly luminous and enshrouded in dusty CSM, our major results differ significantly, in that we find compelling evidence that the internal host-galaxy extinction is quite high, that the star's $T_{\rm eff}$ was cooler (2100--2500 K vs.~2760--3240 K), and, more fundamentally, that the dust composition was oxygen-rich, as we would expect for RSGs, rather than carbon-rich. Why would this one RSG be so inherently different from the hundreds of known RSGs in the Local Group? Ultimately, differences in photometric measurement and approach to the modeling has led to our $L_{\rm bol}$ inference being somewhat higher (up to $\log[L_{\rm bol}/L_{\odot}]=5.19$), although the two values agree to within the uncertainties in each.

Following \citet{Soraisam2023}, it would be intriguing to examine whether our cursory attempt at an estimate of a period of variability for the SN 2025pht progenitor candidate (Section~\ref{sec:variability}) is consistent with the bolometric luminosity we have estimated (Section~\ref{sec:progenitor}). The variability period for RSGs can be empirically correlated with their luminosity (a ``period-luminosity relation''), typically in a given photometric bandpass \citep{Yang2011,Yang2012,Soraisam2018,Ren2019}, such as $K_s$. For the progenitor candidate, the absolute $K_s$ brightness is $M_{K_s}\approx-11.0$ mag (Vega, based on the F200W, F212N and F277W measurements). Recall that our nominal estimates of the star's period are $\sim 470$ and $\sim 660$ days. If we consider the period-luminosity relations from \citet[][LMC]{Yang2011}, \citet[][M31]{Soraisam2018}, and \citet[][M31, M33]{Ren2019}, they would all predict that $M_{K_s}\approx-10.3$ and $\approx-10.8$ mag, respectively. The former is significantly less luminous than the actual value. However, the latter period results in a luminosity that is more consistent with the value we have inferred from the SED modeling. This is reassuring, and tends to indicate that the higher value for the period, $\sim 660$ days, is more likely, if the star was actually an LPV.

The {\tt DUSTY} models provide a good representation of the {\sl JWST\/} observations from 2024. However, they are potentially insufficient at optical wavelengths. Via synthetic photometry, the observed brightness at WFC3 F814W of the best-fitting model would be $\sim 26.8$ mag (Vega), whereas the estimated upper limit in this band is $>25.6$ mag, $\sim 1.2$ mag brighter. (The synthetic brightnesses in the {\sl JWST\/} bands agree with the observations on average by $\sim 0.12$ AB mag.)
The star, however, was observable at F814W in 2001. We had developed in Section~\ref{sec:variability} a plausible, yet grossly oversimplified, pulsational variability model that was consistent with the F814W upper limit in 2024. Possibly the amplitude of the variability was actually more exaggerated than what we inferred with the simple model. Possibly much of the dust shell, the evidence for which was observed in 2024, was set up through enhanced mass loss sometime in the intervening $\sim 23$ yr, and therefore simple, (semi-)regular variability, potentially seen in 2001, no longer applied as it had before. 
%Possibly it never did. Possibly the star had been in its observed state in 2024 for a number of years or decades. We just do not know.
As a possible analog, the optical brightness and periodic variability of WOH G64 had decreased slowly initially and then more dramatically over the last decade or so, together with increased dust formation \citep{Ohnaka2024,vanLoon2026}.

If all we had available to us, photometrically, was the brightest F814W magnitude for the star from 2001 (see Table~\ref{tab:light_curve}), nearest the possible peak of the light curve, and we relied exclusively on a bolometric correction ($BC$), e.g., $BC({\rm F814W})=0.0$ mag \citep{Davies2018}, then we would have arrived at a bolometric magnitude $M_{\rm bol}\approx-7.7$ mag. Assuming $M_{\rm bol}({\odot})=+4.74$ mag \citep{Mamajek2015}, then $\log (L_{\rm bol}/L_{\odot})\approx 5.0$, which is similar to what we have determined here from detailed modeling of the infrared SED of this star. Of course, for all other F814W apparitions in 2001, the star would have been found to be less luminous than this. In fact, if all we had available was the 2024 upper limit, we would have deduced that $\log (L_{\rm bol}/L_{\odot})\lesssim 4.2$! This should therefore be taken as a cautionary warning (not the first time this has been issued; see, e.g., \citealt{Beasor2025}) that the optical, even in the red, poorly exemplifies the actual emitted light from such a star. 

We can obtain an estimate of the mass-loss rate, $\dot M$, from the SN 2025pht progenitor candidate based on the {\tt DUSTY} modeling with {\tt density type = 4}. For the best-fit model, and following \citet[][their Equation 4]{Antoniadis2024}, and assuming Galactic values for the gas-to-dust ratio of 200 and the bulk density of 3 g cm$^{-3}$, and scaling by $L_{\rm bol}$, we found that $\dot M \approx 5.3\times 10^{-5}\ M_{\odot}\ {\rm yr}^{-1}$; if we adopted \citet[][their Equation 6]{vanLoon2007}, we computed a similar $\approx 4.4\times 10^{-5}\ M_{\odot}\ {\rm yr}^{-1}$. This is higher than the rate measured for, e.g.,  Betelgeuse \citep[$\alpha$ Ori; $2.1\times 10^{-7}\ M_{\odot}\ {\rm yr}^{-1}$,][]{DeBeck2010,Dupree2025} and generally somewhat higher than what has been found for Galactic RSGs at the same luminosity as the progenitor candidate \citep[e.g.,][]{Beasor2020,Humphreys2020}, although significantly less than for the extreme RSGs VY CMa in our Galaxy \citep[$6\times 10^{-4}\ M_{\odot}\ {\rm yr}^{-1}$; ][]{Shenoy2016} and WOH G64 in the LMC \citep[$\approx 10^{-3}\ M_{\odot}\ {\rm yr}^{-1}$; ][]{vanLoon1999,vanLoon2005,vanLoon2025}. The short light-curve plateau phase for SN 2025pht may indicate significant depletion of the stellar mantle; this is not surprising for a RSG that had developed, and possibly sustained, mass loss at a high rate.

Although \citet{Kozyreva2025} called for a precursor event for SN 2023ixf, roughly analogous to what \citet{Davies2022} proposed --- though observational evidence has not supplied support for such an event in this case \citep[e.g.,][]{Dong2023} --- the situation for the SN 2025pht progenitor candidate is comparatively unconstrained. We are not aware of any other available pre-SN data with sufficient spatial resolution that might provide us with clues as to the star's nature during that decades-long time interval between 2001 and 2024.

\begin{figure}
\plotone{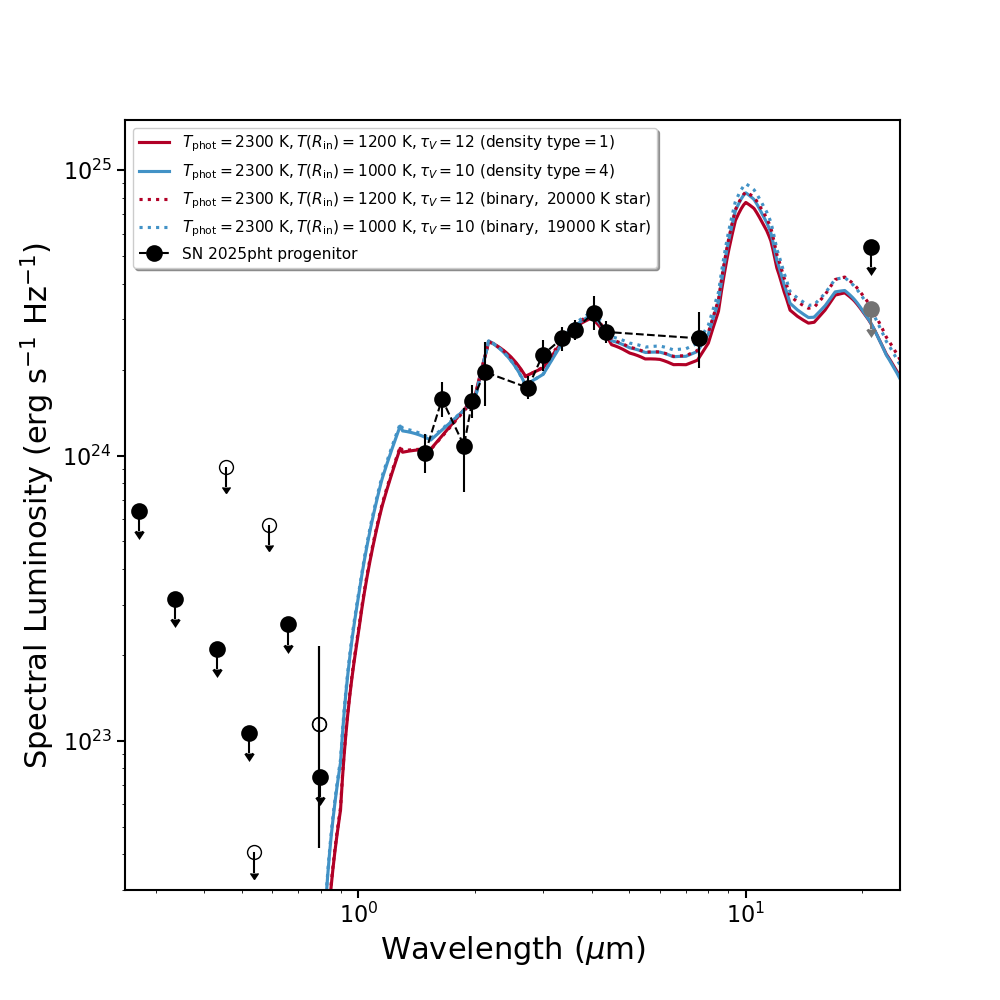}
\caption{The same as for Figure~\ref{fig:prog_fit}, however, with just the best-fitting models assuming {\tt density type=1} and {\tt density type=4}. Here we have also added a hot 20,000 K model photosphere with luminosity $2000\ L_{\odot}$ (approximately of B2 spectral type; \citealt{Castelli2003}) to the cool {\tt PHOENIX} model photosphere \citep{Kucinskas2005,Kucinskas2006} at 2300 K, as input to {\tt DUSTY} for {\tt density type=1}; and, a 19,000 K photosphere with $1500\ L_{\odot}$  (approximately B3-type) for {\tt density type=4}. Both of these binary models describe the observed SED reasonably well and do not violate the {\tt space{\textunderscore}phot}-estimated upper limit in {\sl JWST\/} MIRI F2100W. Any companion model hotter, and therefore more luminous, than these assumed B-type stars can be excluded.
\label{fig:binary}}
\end{figure}

Furthermore, the interplay between pulsations and interactions with a binary companion could affect the pre-SN envelope and CSM structure (\citealt{Laplace2026} and references therein; see also \citealt{vanLoon2026} with respect to WOH G64). %However, the uncertainties that we have needed to place on the ensemble of SED measurements for the progenitor candidate preclude us here from a meaningful constraint on a companion star \citep[see, e.g.,][in the case of SN 2023ixf]{VanDyk2024}.
Following \citet{VanDyk2024} we placed constraints on the presence of a companion by including a hot stellar photosphere together with the cool photosphere, scaling by relative luminosity, as the input source to {\tt DUSTY}. We performed this re-run of the best-fit dust models, assuming both {\tt density type=1} and {\tt density type=4}. For the former we constrained a putative companion to a star with $T_{\rm eff}\lesssim20,000$ K and $L_{\rm bol}\lesssim2,000\ L_{\odot}$, and for the latter, a similar $\lesssim19,000$ K and $\lesssim1,500\ L_{\odot}$, both limits consistent with a star of spectral type B2 to B3; see Figure~\ref{fig:binary}.

It should now go without saying that the candidate needs to be confirmed as the actual progenitor (e.g., \citealt{VanDyk2023} and references therein), through, say, future {\sl JWST\/} observations to determine whether the star has disappeared. This could be challenging, though, since we are expecting excess late-time emission from long-term CSM interaction. Additionally, we expect dust formation for SNe showing evidence for interaction \citep[e.g.,][]{Sarangi2015,Sarangi2022}, which will manifest itself with increased flux in {\sl JWST\/} bands and thwart efforts to determine that the candidate has vanished.

This star, and other stars similar to it, could possibly go completely undetected in bulk searches for RSGs in galaxies. \citet{Sarbadhicary2026} recently completed a study in which they isolated nearly 100,000 candidate RSGs across several nearby galaxies. Their selection algorithm consisted of a {\sl HST}+{\sl JWST\/} color and luminosity ``wedge.'' The SN 2025pht progenitor candidate had color $({\rm F814W}-{\rm F200W})_0\gtrsim3.5$ mag and luminosity $M_{\rm F200W}\approx-8.7$ mag, which would have placed the star not only outside of the wedge boundaries, but also potentially beyond the nominal detection limits of their galaxy survey, at least in this filter combination. Complicating detection further, of course, was the significant host-galaxy reddening in the case of SN 2025pht. A similar example of a star that would likely have been missed in previous surveys of galaxies beyond the Local Group is the extreme RSG WOH G64, which has attracted considerable attention from its unexpected behavior over the past two decades \citep{Levesque2009,Ohnaka2024,vanLoon2026}. Possibly other color spaces, based on {\sl JWST\/} bands alone, would provide better means of selecting these especially dusty and luminous RSGs.

In the case of SN 2025pht we have been extraordinarily fortunate that we have photometry with exceptional wavelength coverage, especially in the near- to mid-infrared, of the progenitor candidate, which are contemporaneous, or nearly-contemporaneous --- what we earlier termed a ``quasi-snapshot,'' which reveals the star's SED in extraordinary detail. This is astoundingly rare, to date, although progressively more {\sl JWST\/} programs are covering a number of nearby potential SN host galaxies in various bands. Though we should appreciate the current stroke of luck, it is possible that this could occur again soon. However, what could well be lacking is sampling of the star's variability. We know that RSGs can be LPVs and, in fact, that their colors vary with their pulsations (e.g., \citealt{Ogane2022} in the particular case of Betelgeuse; the SN 2023ixf progenitor candidate, for instance, varied in $H-K$ by $\sim 0.68$ mag over just $\sim 34$ days, \citealt{Soraisam2023}). Possibly, archival {\sl HST\/} data at one or more epochs would also exist, as in the current case. Potentially, through a combination of time-series coverage of nearby hosts via the Vera C.~Rubin Observatory, together with the superb sensitivity and spatial resolution of {\sl JWST}, and the wide-field, near-infrared coverage of the upcoming {\sl Nancy Grace Roman Space Telescope}, the community may be able to gather more meaningful and instructive statistics on the nature of SN II-P progenitors.

\begin{acknowledgements}
We are grateful to the reviewer for their careful reading of the manuscript and helpful recommendations toward its improvement, particularly with regards to the distance estimate for the host galaxy and the overall {\tt DUSTY} modeling of the progenitor candidate SED.
We thank Thomas de Jaeger for pointing us to the SN 2008M photometry of \citet{Anderson2024}. No artificial intelligence was used to create and write this manuscript. This research is based in part on observations made with the NASA/ESA Hubble Space Telescope, obtained from the Space Telescope Science Institute (STScI), which is operated by the Association of Universities for Research in Astronomy (AURA), Inc., under NASA contract NAS 5–26555. These observations are associated with programs GO-9042, GO-9155, GO-5446, and GO-17502. This work is based in part on observations made with the NASA/ESA/CSA James Webb Space Telescope. The data were obtained from the Mikulski Archive for Space Telescopes (MAST) at STScI, which is operated by AURA, Inc., under NASA contract NAS 5-03127 for JWST. These observations are associated with programs GO-3707 and GO-4793. A.V.F.'s research group at U.C. Berkeley was supported in part by NASA/JWST grants GO-03921 and AR-06356 from STScI, which is operated by AURA, Inc., under NASA contract NAS5-26555. Additional financial support was provided by Gary and Cynthia Bengier, Clark and Sharon Winslow, Alan Eustace and Kathy Kwan (W.Z. is a Bengier-Winslow-Eustace Specialist in Astronomy), Timothy and Melissa Draper, Briggs and Kathleen Wood, Ellyn and Alan Seelenfreund (T.G.B. is Draper-Wood-Seelenfreund Specialist in Astronomy), and numerous other donors. A major upgrade of the Kast spectrograph on the Shane 3\,m telescope at Lick Observatory, led by Brad Holden, was made possible through gifts from the Heising-Simons Foundation, William and Marina Kast, and the University of California Observatories. Research at Lick Observatory is partially supported by a gift from Google. BHTOM.space is based on the open-source TOM Toolkit by LCO and has been supported by the European Union's research and innovation programmes under grant agreements 101004719 (OPTICON-RadioNet Pilot, ORP) and 101131928 (ACME). The BHTOM project has received funding from the European Union's Horizon Europe Research and Innovation programme ACME under grant agreement 101131928 (2024–2028). {\L}.W. acknowledges support from the Polish National Science Centre DAINA grant 2024/52/L/ST9/00210. AR acknowledges financial support from the SOXS project (PI S.~Campana). Partially based on observations collected at the Schmidt telescope (Asiago Ekar, Italy) of the INAF–Osservatorio Astronomico di Padova.
\end{acknowledgements}

\begin{contribution}
S.D.VD undertook much of the analysis and led construction of the manuscript. T.S. and N.Z. both provided {\tt space{\textunderscore}phot} photometry of the {\sl JWST\/} data. G.S.A. provided the JAGB host-galaxy distance analysis. T.G.B. conducted the Lick/Kast observations (PI A.V.F.) of SN 2025pht and reduced the  spectra; W.K.Z. observed on 2025 August 30. L.W., P.J.M., K.K., F.J.H., A.W., M.Z., and A.R. were responsible for obtaining and distributing online the optical SN 2025pht light curves. D.M., O.D.F., J.E.J., W.Z., A.V.F., and all other authors supplied cogent and valuable input, and provided detailed comments and edits on the original draft version.
\end{contribution}

\facilities{HST (WFPC2, WFC3), JWST (NIRCam, MIRI), Shane (Kast Double spectrograph), BHTOM}

\software{
          {\tt AstroDrizzle} \citep{STScI2012b},
          {\tt Dolphot} \citep{Dolphin2016},
          {\tt DUSTY} \citep{Ivezic1999},
          {\tt extinction} \citep{Barbary2016},
          {\tt IRAF} \citep{Tody1986}, 
          {\tt iraf{\textunderscore}docker} (https://github.com/tepickering/iraf{\textunderscore}docker),
          {\tt IDLAstro} \citep{Landsman1995},
          {\tt PyRAF} \citep{STScI2012a},
          {\tt pysynphot\/} \citep{STScI2013},
          {\tt space{\textunderscore}phot} \citep{Pierel2024}
          }

\appendix

\section{Estimations of Reddening to the SN}

\subsection{Reddening From Interstellar Na~{\sc i}~D}\label{sec:naid}

\begin{figure}
\plotone{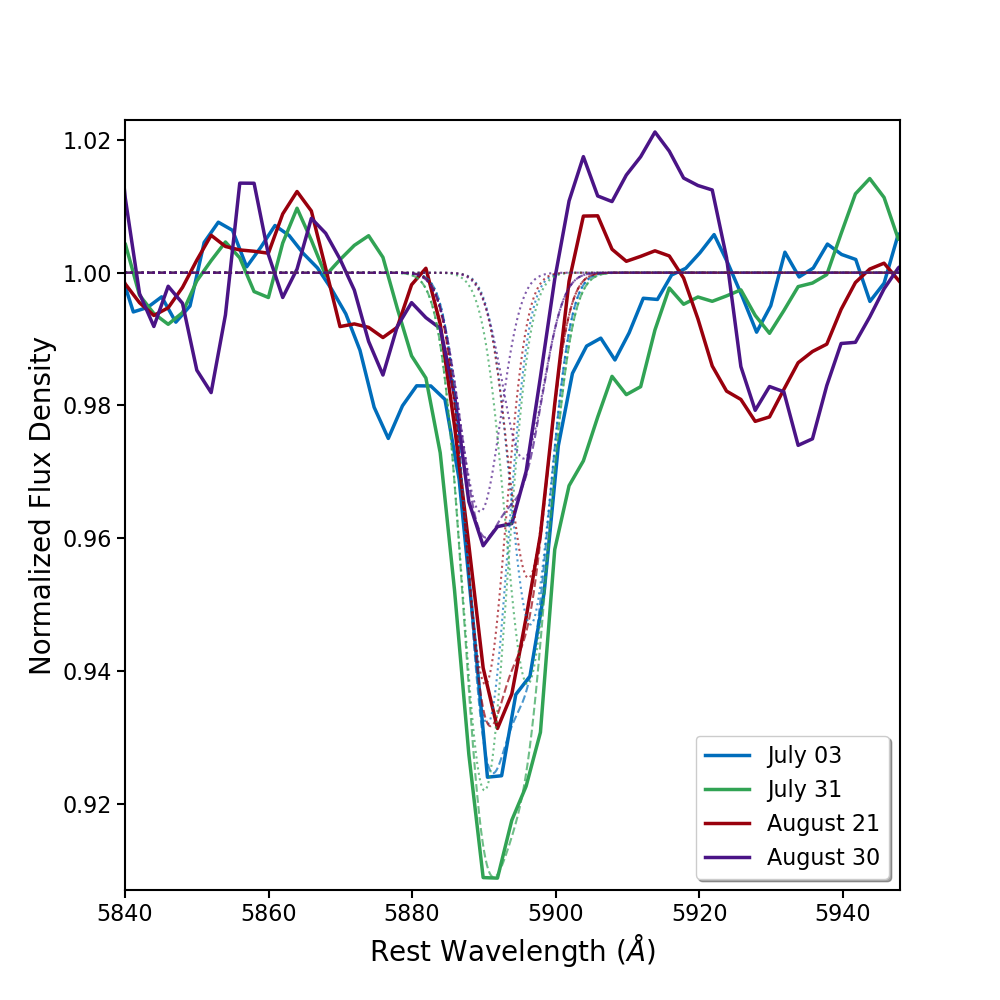}
\caption{The region around the Na~{\sc i}~D feature in flux-normalized optical spectra (solid curves) of SN 2025pht from 2025 July 3 \citep{Strader2025} and from July 31, August 21, and August 30 (all in this study). The spectra have  been corrected for the redshift of the host galaxy. Also shown are simple Gaussian models at the individual rest wavelengths of the D1 and D2 lines (dotted curves), of which the feature is comprised, as well as the sum of these two models (dashed curves).}
\label{fig:naid}
\end{figure}

We show the Na~{\sc i}~D absorption, as seen in each of the four SN 2025pht spectra, in Figure~\ref{fig:naid}. Contrary to the claim made by \citet{Kilpatrick2025}, it is evident that the host-galaxy Na~{\sc i}~D feature is clearly detectable not only in the classification spectrum, but also in the later spectra. This feature had not vanished.

We do note, though, that the total EQW of the feature appears to vary with time. We fit the total EQW for each spectrum with a combination of two simple Gaussians which approximate the two components of the feature, D1 at 5895.92 \AA\ and D2 at 5889.95 \AA\ (these two components are unresolved in all of these spectra). For the July 3 spectrum, from the model we find EQW $\approx 0.86$ \AA; July 31, $\sim 1.09$ \AA;  August 21, $\sim 0.77$ \AA; and August 30, $\sim 0.46$ \AA.  We do not infer that the change in EQW implies that the reddening has also varied with time; the evidence from the other methods, presented below, argues against this being the case. The varying EQW might arise from differences in instrumental setup, observing, and spectral extraction. However, more likely is the fact that Na~{\sc i}~D from the SN itself emerges starting at $\sim 30$ days and develops a strong P-Cygni profile \citep{Gutierrez2017}. One can see in Figure~\ref{fig:spectracomp} this feature strengthening in the later SN 2025pht spectra. Such SN emission evolving with time in this wavelength region could fill and dilute, at least partially, the presence and strength of the interstellar feature.

Thus, the use of the Na~{\sc i}~D EQW as a measure of $A_V$ should be limited (at best) to when an SN is quite young (a few days in age),  when the spectrum is essentially a featureless continuum (modulo the presence of any ``flash ionization'' emission lines).

\subsection{Reddening from Spectral Comparison}\label{sec:spectra}

\begin{figure}
\gridline{\fig{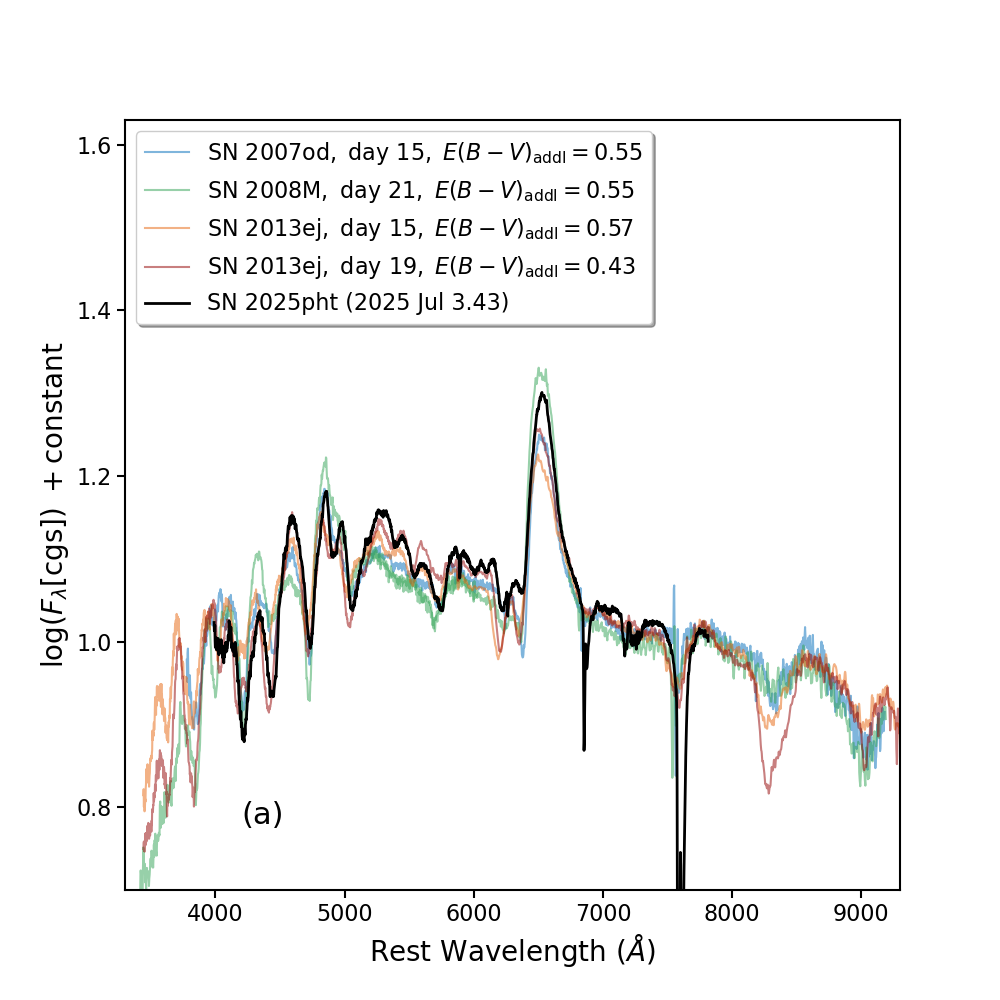}{0.48\textwidth}{}
          \fig{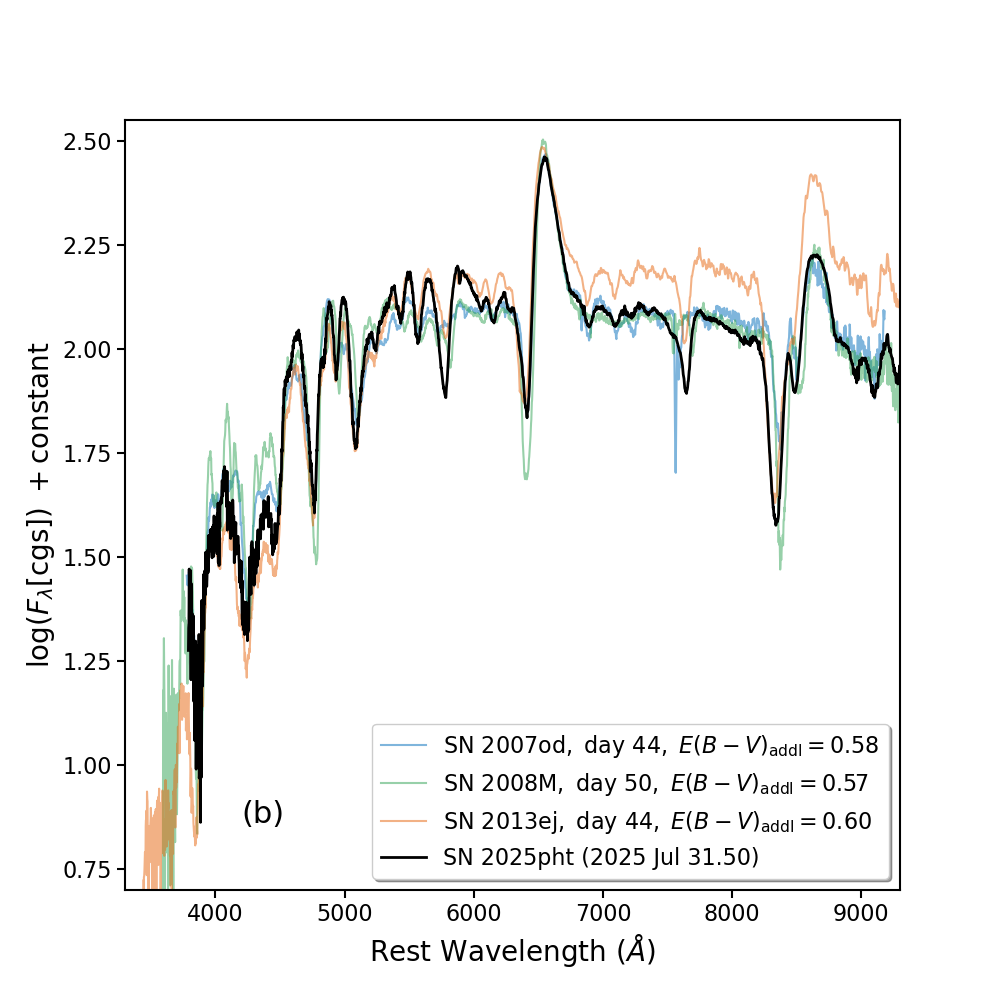}{0.48\textwidth}{}}
\gridline{\fig{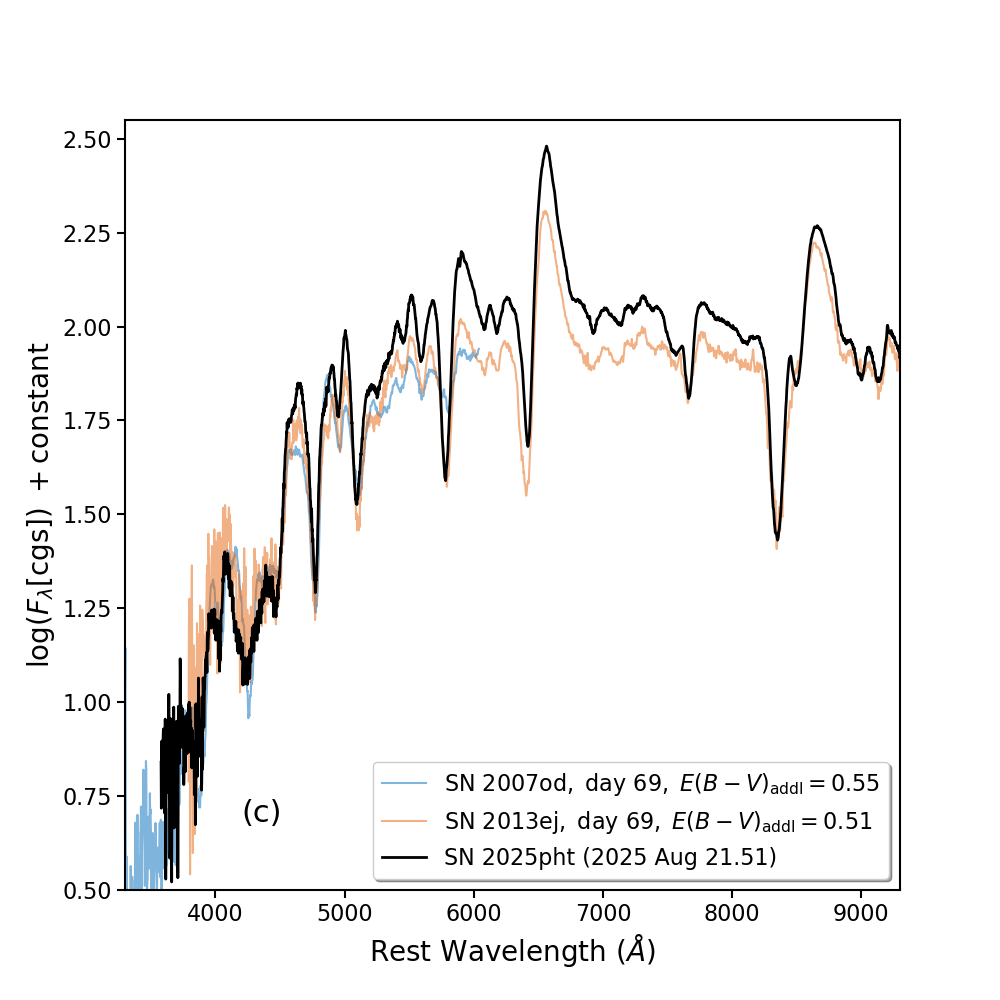}{0.48\textwidth}{}
          \fig{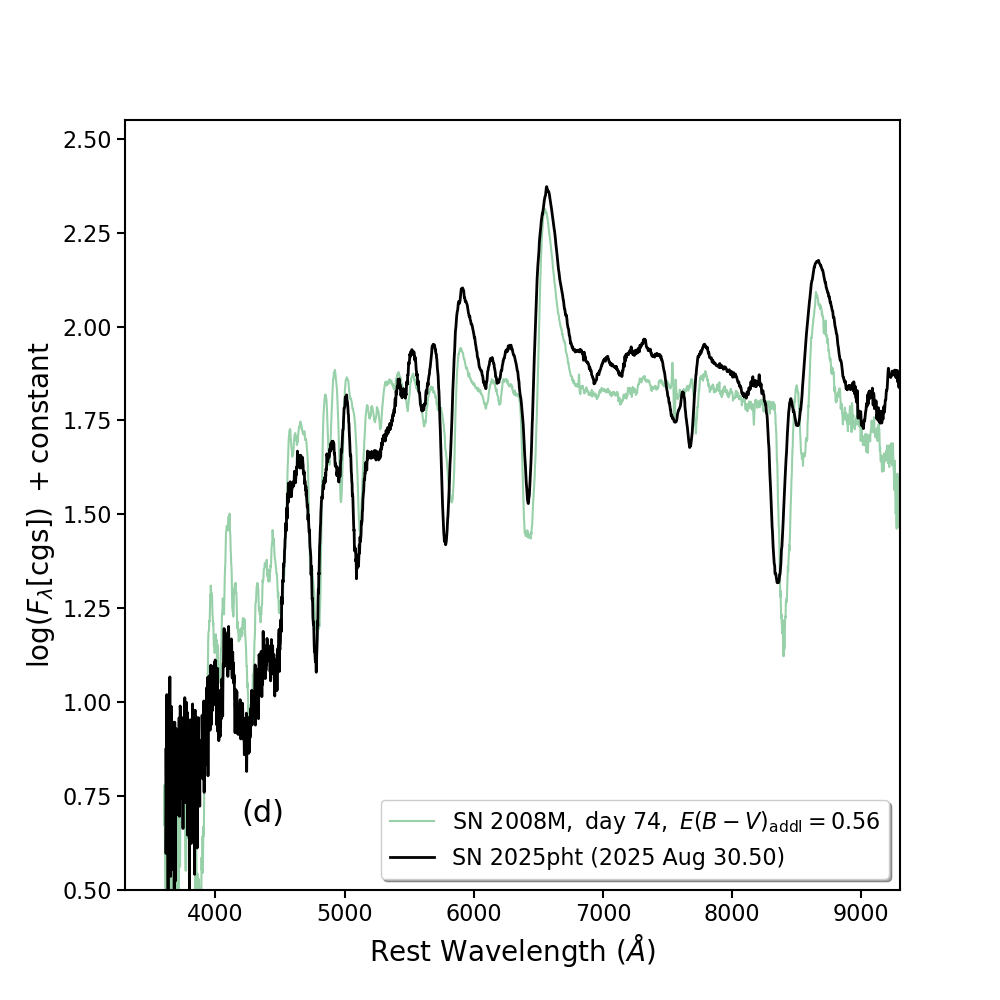}{0.48\textwidth}{}}
\caption{Optical spectra of SN 2025pht. We show the SN 2025pht classification spectrum from 2025 July 3 \citep{Strader2025} in panel (a), and spectra from 2025 July 31, August 21, and August 30 (all from this study) in panels (b), (c), and (d), respectively. Displayed for comparison in the panels are spectra of SN 2007od and SN 2008M \citep{Gutierrez2017}, as well as of SN 2013ej \citep{Dhungana2016}, at ages indicated in the panel legends. The spectra have been corrected for host-galaxy redshift, as well as for Galactic foreground reddening \citep{Schlafly2011}. Both SN 2007od and SN 2008M have effectively zero internal host-galaxy extinction ($A_V[{\rm host}]\approx0$ mag; \citealt{Anderson2014}); however, we have further corrected SN 2013ej for host reddening $E(B-V)_{\rm host}\approx0.07$ mag, following \citet{Huang2015}. The comparison spectra were then artificially reddened by additional amounts, $E(B-V)_{\rm addl}$ (as indicated in the panel legends), and fit to the SN 2025pht spectra.}
\label{fig:spectra}
\end{figure}

We show in Figure~\ref{fig:spectra} the spectra of SN 2025pht. \citet{Anderson2014} concluded that SN 2007od did not experience host reddening ($A_V=0.00\pm0.06$ mag). We therefore considered spectra of SN 2007od \citep{Gutierrez2017} for comparison.  Additionally, we considered spectra of SN 2008M (also \citealt{Gutierrez2017}), which was also free of host extinction ($A_V=0.00\pm0.07$ mag; \citealt{Anderson2014}), as well as of SN 2013ej, for which we further corrected for host reddening, $E(B-V)_{\rm host}\approx0.07$ mag, following \citet{Huang2015}.

We subsequently artificially reddened the comparison SN spectra until their overall shapes matched the SN 2025pht spectra. Here we used {\tt pysynphot\/} \citep{STScI2013} first to deredden all of the spectra by their respective Galactic foreground components (\citealt{Schlafly2011}; the \citealt{Cardelli1989} reddening law is assumed). If necessary, we further dereddened for the host-galaxy component. This only may have applied to SN 2013ej: most of the studies of this event assumed that the host-galaxy reddening was negligible, but \citet{Huang2015} estimated that $E(B-V)_{\rm host} = 0.07\pm0.08$ mag, which we have assumed here. (In fact, we argue in Appendix~\ref{sec:color_comp} that even this amount was likely an underestimate of the possible host reddening for this SN.) All of the spectra were then corrected again for host redshift. We then reddened the comparison SNe with {\tt pysynphot\/} by amounts of additional reddening, $E(B-V)_{\rm addl}$, as shown in the Figure~\ref{fig:spectra} legends, and fit them to the SN 2025pht spectra.

\subsection{$B-V$ Color Comparison}\label{sec:bminv}

\begin{figure}
\plotone{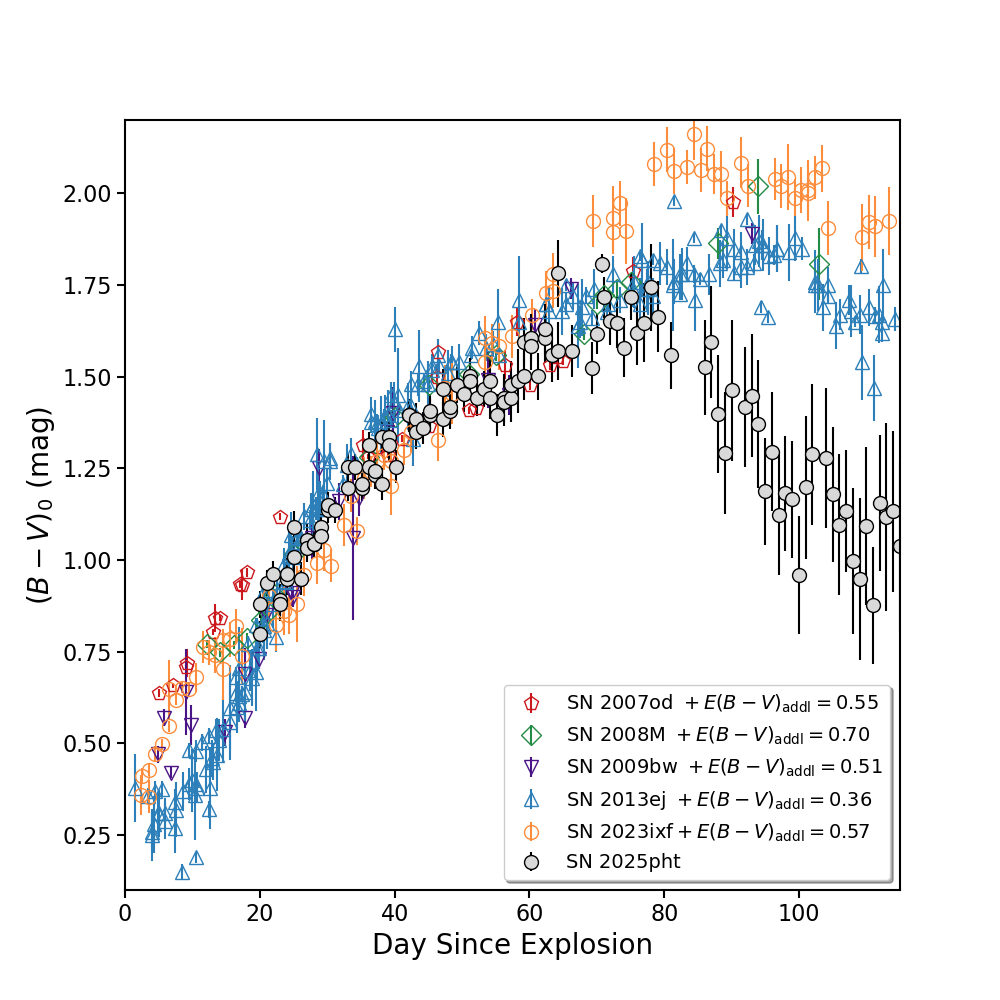}
\caption{Optical $B-V$ color evolution of SN 2025pht, based on BHTOM photometry \citep{Mikolajczyk2025}. Also shown for comparison are $B-V$ color curves for SN 2007od \citep{Inserra2011,Anderson2024}, SN 2008M \citep{Anderson2024}, SN 2009bw \citep{Inserra2012}, SN 2013ej \citep{Valenti2014,Bose2015,Huang2015,Dhungana2016,Yuan2016}, and SN 2023ixf \citep{Zheng2025}. All of the SN color curves were initially corrected for Galactic foreground reddening \citep{Schlafly2011}. All of the comparison SNe were further corrected for any host-galaxy reddening. The quantity $E(B-V)_{\rm addl}$ represents the amount of additional reddening that was required for a comparison SN color curve to fit the SN 2025pht curve; see text.}
\label{fig:bminv}
\end{figure}

We show the SN 2025pht $B-V$ color in Figure~\ref{fig:bminv}.
From a fit, limited to the range days 21--69, we found for SN~2007od that $E(B-V)_{\rm addl}=0.55$ mag, and for SN 2009bw a similar $E(B-V)_{\rm addl}=0.51$ mag. The fit for SN~2023ixf required an additional 0.57 mag of reddening. The amount of necessary $E(B-V)_{\rm addl}$ for SN 2008M appears to be larger; however, as we point out in Section~\ref{sec:color_comp}, this SN is in the sample of intrinsically bluest SNe from \citet{deJaeger2018}, and that sample can be as much as $\sim 0.23$ mag bluer than the \citet{deJaeger2018} sample with no $A_V({\rm host})$. The fit for the SN 2013ej color curve indicates comparatively less required $E(B-V)_{\rm addl}$, although, again from Appendix~\ref{sec:color_comp}, we speculate that the host reddening component for that SN could have been underestimated by $E(B-V)_{\rm host}\approx0.10$--0.25 mag (significantly more than assumed by the various studies of the SN). We note that the abrupt change from a progressively redder $B-V$ to a blue $B-V$ at the end of the plateau is most pronounced for the short-plateau SN 2025pht.

\subsection{A Further Look at the $B-V$ Color Comparison}\label{sec:color_comp}

\begin{figure}
\plotone{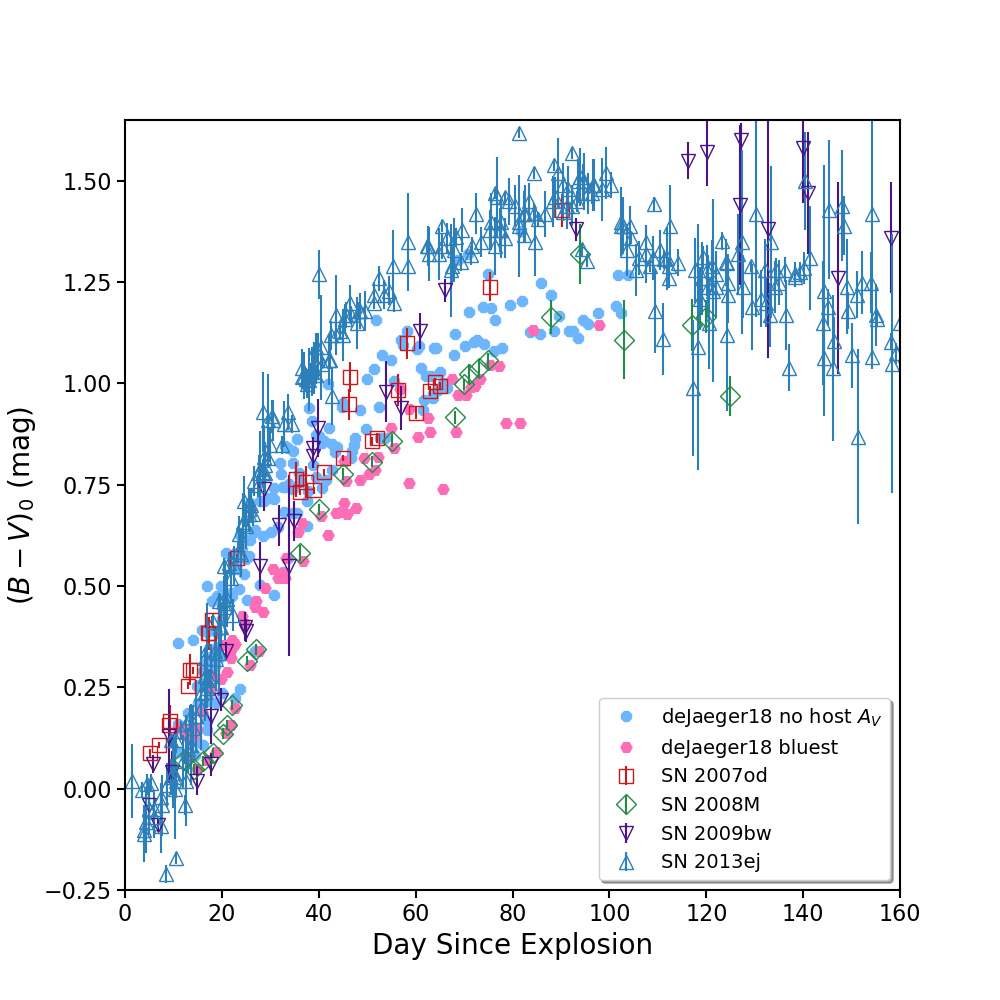}
\caption{$B-V$ color curves for SN 2007od \citep{Inserra2011,Anderson2024}, SN 2008M \citep{Anderson2024}, SN 2009bw \citep{Inserra2012}, and SN 2013ej \citep{Valenti2014,Bose2015,Huang2015,Dhungana2016,Yuan2016}. The curves for all of these SNe have been corrected for Galactic foreground reddening \citep{Schlafly2011}. Also shown for comparison are the SN samples from \citet{deJaeger2018} (a) with no apparent internal host extinction (``no host $A_V$'') and (b) which are intrinsically blue (``bluest'').
\label{fig:bv}}
\end{figure}

We have compared the $B-V$ color curves, after correction for Galactic foreground reddening, of four of the comparison SNe that we have considered: SN 2007od \citep{Inserra2011,Anderson2024}, SN 2008M \citep{Anderson2024}, SN 2009bw \citep{Inserra2012}, and SN 2013ej \citep{Valenti2014,Bose2015,Huang2015,Dhungana2016,Yuan2016}; see Figure~\ref{fig:bv}. We have also included the color curves for the sample of SNe with no apparent host extinction, $A_V{\rm (host)}$, and those with the intrinsically bluest colors from \citet{deJaeger2018}. We confirmed that SN 2007od appears to have essentially no host extinction, as reported by \citet[][$A_V({\rm host})=0.00\pm0.06$ mag]{Anderson2014}. Similarly, SN 2008M shows no evidence for host extinction ($A_V({\rm host})=0.00\pm0.07$ mag; \citealt{Anderson2014}), however, it also appears to have been intrinsically very blue and, for that reason, may not serve as an adequate color comparison for SN 2025pht. The \citet{deJaeger2018} sample with no $A_V{\rm (host)}$ is $\sim 0.23$ mag redder at $\sim 50$ days than the intrinsically bluest sample (although the dispersion across the former sample is $\sim 0.11$ mag at that age). SN 2009bw appears to be fairly blue and also consistent with essentially no $A_V({\rm host})$ (\citealt{Inserra2012} estimated that $E[B-V]_{\rm host}$ was low, $\sim0.08$ mag). Finally, the color of SN 2013ej is just along the reddest edge of the colors for the SNe with no $A_V({\rm host})$ and, in fact, we found that, although it was somewhat unusually blue for the first $\sim20$ days, the color curve could be dereddened even further, by $E(B-V)\approx0.10$--0.25 mag, to better agree with the $A_V({\rm host})=0$ mag sample. Authors that studied SN 2013ej photometrically generally assumed that $A_V({\rm host})\approx0$ mag, except for \citet{Huang2015}, who estimated that either $E(B-V)_{\rm host}\approx 0.06\pm0.06$ or $0.07\pm0.08$ mag, depending on method, both of which, of course, are also consistent with zero host reddening.

\section{Host-Galaxy Distance Estimations}\label{sec:distance_estimates}

\subsection{Another SCM Distance Estimate for SN 1999em}\label{sec:SCM_99em}

Following \citet{Polshaw2015}, the SCM is based on the $I$-band brightness and the Fe~{\sc ii} expansion velocity at age 50 days. It appears that the SCM described by \citet{Olivares2008} was based on 30 days. We estimated the explosion epoch from the \citet{Jones2009} EPM value, for which those authors found $t_0$ = JD 2,451,476.3 $\pm 1.1$ and 2,451,474.0 $\pm 2.0$ using the E96 and D05 models, respectively. These values for the explosion epoch agree to within the uncertainties;  taking the average, 
$t_0$ = 2,451,475.2, MJD 51474.7, which is 1999 Oct 23.7. This agrees with what \citet{Elmhamdi2003} had found, 1999 Oct 24.5. Adding 50 days to this yields MJD 51524.7 (1999 Dec 12.7).

We estimated that the $I$ magnitude on day 50 is $I_{50}=13.29 \pm 0.03$, which is a weighted average of the photometry around this date from \citet{Hamuy2001}, \citet{Leonard2002}, and \citet{Elmhamdi2003}. We obtained the \citet{Leonard2002} spectrum of SN 1999em from 1999 December 12 from WISeREP. After correcting the spectrum for the host-galaxy redshift, we measured the velocity at Fe~{\sc ii} $\lambda$5169 on day 50, $v_{50}(5169)$, to be $3810 \pm 270$ km s$^{-1}$. We also estimated a photospheric velocity, $v_{50}({\rm phot})$, which is the weighted average of $v_{50}(5169)$ together with the velocities at Fe~{\sc ii} $\lambda$$\lambda$4629, 4924, and 5018: a similar $3620 \pm 120$ km s$^{-1}$.

The remaining quantity to be determined is the total extinction $A_I({\rm tot})$. \citet{Jones2009} assumed that the host-galaxy visual extinction is $A_V=0.24\pm0.14$ mag, whereas \citet{Olivares2008}, from \citet{Dessart2006}, assumed $A_V=0.31\pm0.16$ mag. The weighted average of these two is $A_V=0.27\pm0.15$ mag and, assuming $R_V=3.1$, $A_I({\rm host})=0.17\pm0.09$ mag. Including the Galactic foreground $A_I({\rm Gal})=0.061$ mag \citep{Schlafly2011} leads to $A_I({\rm tot})=0.23\pm0.09$ mag.

Based on $v_{50}(5169)$ we found that $d=10.45\pm2.32$ Mpc, and 
from $v_{50}({\rm phot})$, $d=9.86\pm1.82$ Mpc, which agree with each other to within the (large) uncertainties. These values also agree with the SCM distance by \citet{Olivares2008}, again, to within the uncertainties. At the very least, we have provided a cross-check on the previous estimate, employing a different SCM calibration. Note that \citet{Polshaw2015} based their calibration of SCM on a value of $H_0=73.8$ km s$^{-1}$ Mpc$^{-1}$ from \citet{Riess2011}, whereas higher-precision, and somewhat reduced, values of $H_0$ have been presented more recently \citep[e.g.,][]{Riess2016,Riess2022}.

\subsection{An SCM Distance Estimate from SN 2025pht}\label{sec:SCM_25pht}

\begin{figure}
\plotone{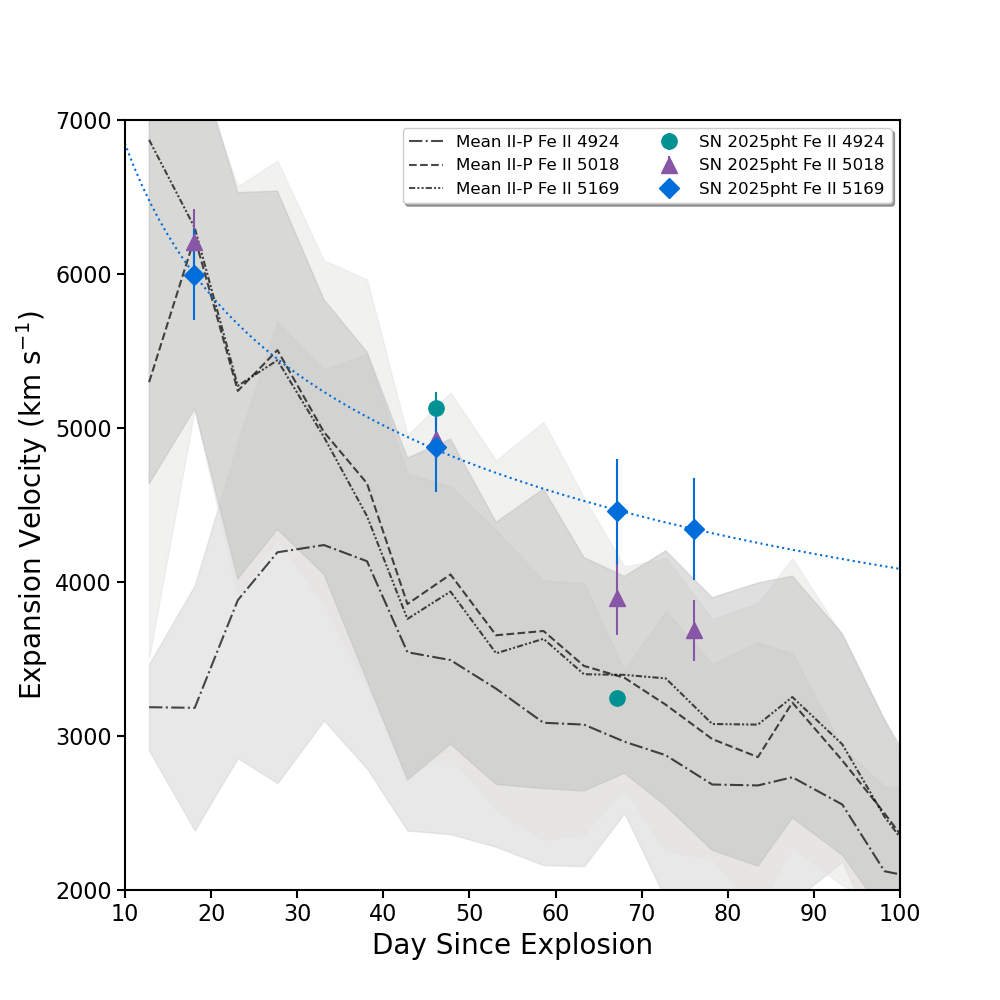}
\caption{Expansion velocities for SN 2025pht, based on the Fe~{\sc ii} absorption lines. Shown for comparison are the mean velocities, and their uncertainties, for the SN II-P sample from \citet{Gutierrez2017}.}
\label{fig:velocities}
\end{figure}

We measured expansion velocities from the various Fe~{\sc ii} lines in the four spectra of SN 2025pht that we consider. Our measurements are shown in Figure~\ref{fig:velocities}, together with the comparison trends from \citet{Gutierrez2017}. We were able to measure velocities for both the Fe~{\sc ii} $\lambda$5018 and $\lambda$5169 lines in all four spectra; however, the $\lambda$4924 line did not become apparent until the July 31 spectrum. As can be seen, the velocities, particularly from the $\lambda$5169 line, are on the relatively high side, compared to the \citet{Gutierrez2017} sample.

From the BHTOM photometry \citep{Mikolajczyk2025}, measurements in $I$ are available on day 47 ($13.15\pm0.06$ mag) and on day 53 ($13.20\pm0.01$ mag), but not on day 50 itself. Instead, we interpolated from these a value of $I_{50}=13.21\pm0.05$ mag. Based on our estimate of $A_V({\rm tot})$, the total $I$ extinction is $A_I({\rm tot})=1.12\pm0.16$ mag. We have chosen here to base the SCM distance only on $v_{50}(5169)$ ($=4770\pm310$ km s$^{-1}$), since we were unable to measure velocities consistently for the other Fe~{\sc ii} lines, as we have pointed out above. Ultimately, we arrived at a distance of $8.65\pm0.88$ Mpc. Even though the velocities are high for typical SNe II-P (the velocity trend here may be consistent with a higher initial mass, i.e., $M_{\rm ini}\gtrsim15\ M_{\odot}$; \citealt{Laplace2026}), the main influence on the value of the distance computation is the total extinction, which we know for SN 2025pht (from Section~\ref{sec:reddening}) is quite high. 

\subsection{A JAGB Distance For the Host Galaxy}\label{sec:jagb}

Following \citet{Li2024,Li2025b}, we considered {\tt Dolphot} photometry across all chips in F150W and F277W, using the ``warmstart'' option with the F150W output as the input star list for the F277W photometry. We subsequently filtered the merged star list in both bands with the following photometric quality cuts for the {\tt Dolphot} output parameters: crowding $<0.5$, sharpness$^2$ $\le 0.01$, object type $\le 2$, signal-to-noise ratio $S/N\ge5$, and photometry quality flag $\le 2$ \citep{Warfield2023, Anand2024}. We then corrected both bands for Galactic foreground extinction. %The resulting color-magnitude diagram (CMD) is shown in Figure~\ref{fig:jagb_cmd}.

\begin{figure}
\plotone{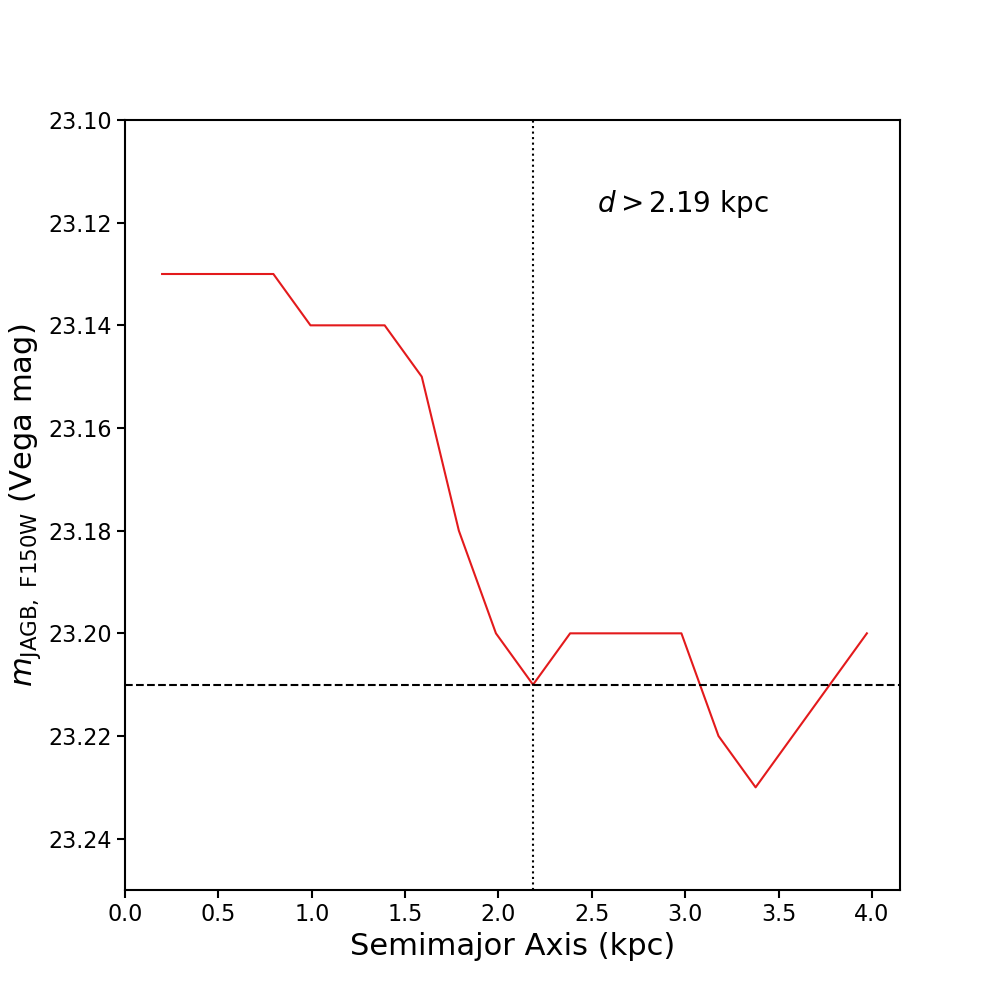}
\caption{The JAGB F150W brightness, corrected for Galactic foreground extinction \citep{Schlafly2011}, as a function of radial distance from the NGC 1637 nucleus (solid red curve; following \citealt{Lee2025}). Each data point is the mode measured from the JAGB luminosity function. The dotted line indicates the radial distance at which the JAGB brightness initially ``converges'' to the level indicated by the dashed line, which in this case is $m_{{\rm JAGB},\ F150W}=23.21$ mag; see Figure~\ref{fig:jagb_cmd}.
\label{fig:ellipses}}
\end{figure}

Next, following \citet{Lee2025}, we applied a series of concentric elliptical masks (at a position angle of 20\arcdeg), centered on the galactic nucleus, with progressively increasing major and minor axes, to mitigate against the effects of crowding and interstellar dust in the inner parts of the host, and establish a convergence in the mode brightness of the JAGB. We isolated the JAGB in the CMD as stars in the color range $1.0 \le ({\rm F150W}-{\rm F277W})_0\le1.5$ mag \citep{Li2024}. In Figure~\ref{fig:ellipses} we display this modal brightness at F150W as a function of semimajor axis.

\begin{figure}
\plotone{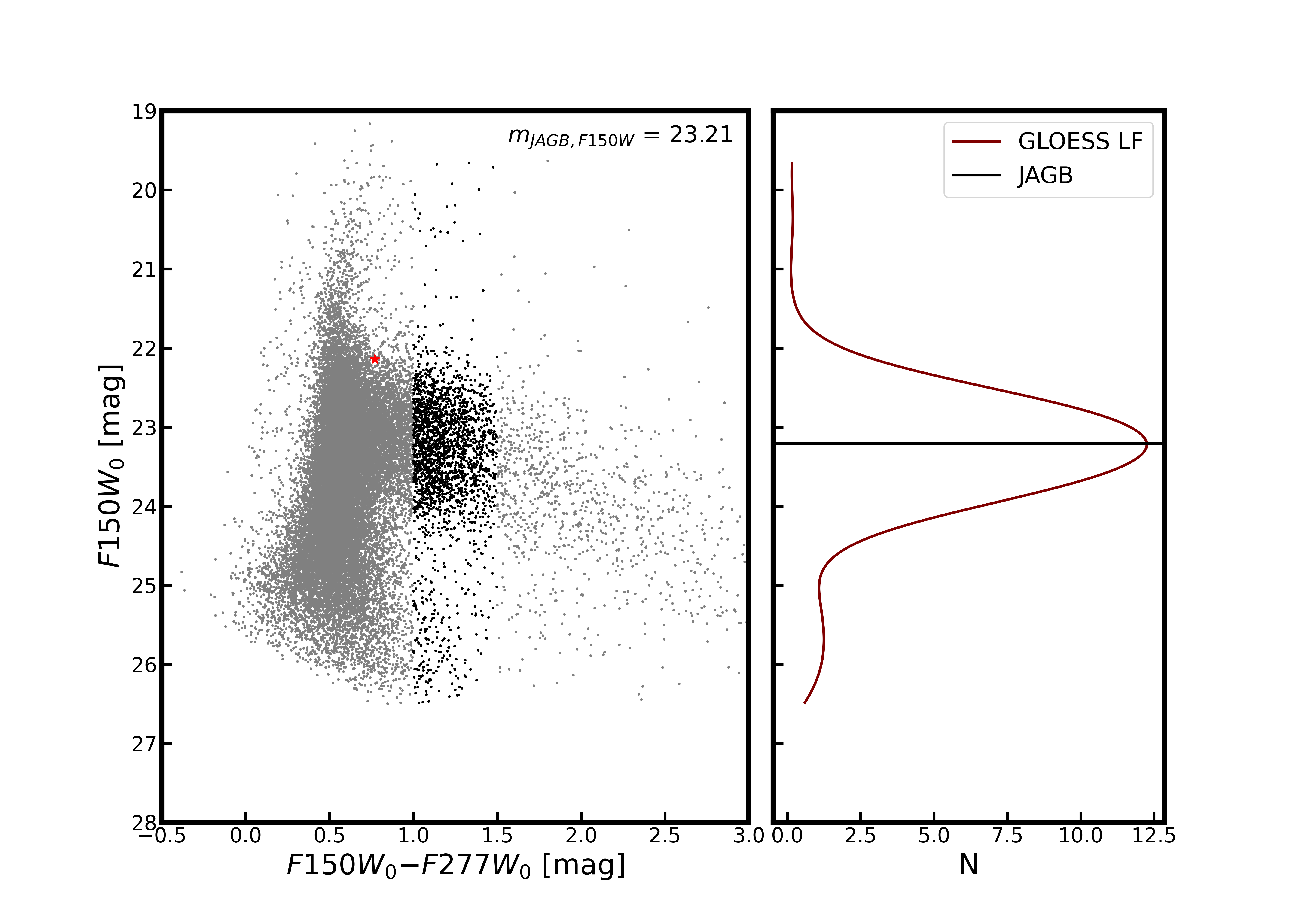}
\caption{{\it Left:} CMD in F150W and F277W for the stars in NGC 1637 beyond a radial distance of 2.19 kpc from the galactic nucleus; see Figure~\ref{fig:ellipses}. The data have been corrected for Galactic foreground extinction to the host galaxy \citep{Schlafly2011}. The black points are those stars within the color range isolating the J-region used in the JAGB estimation, i.e., $1.0\le (F150W_0-F277W_0)\le 1.5$ mag, following \citet{Li2024,Li2025b}. The red star indicates the locus of the progenitor candidate in these bands, for reference. {\it Right:} J-region luminosity function. The histogram for these data (not shown) were binned by widths of 0.01 mag; we then applied a Gaussian-windowed, locally weighted scatterplot smoothing (GLOESS) with a smoothing parameter $s=0.25$ mag. The smoothed luminosity function is shown as the red curve. The solid line indicates the peak of the function, which we adopt as the JAGB brightness, $m_{JAGB,\ F150W}=23.21$ mag.
\label{fig:jagb_cmd}}
\end{figure}

We show in Figure~\ref{fig:jagb_cmd} (left panel) the resulting color-magnitude diagram (CMD) at the first radius at which the mode appears to converge (at $\sim2.19$ kpc). The right panel illustrates a histogram, with a Gaussian-windowed, Locally Weighted Scatterplot Smoothing  \citep[GLOESS; e.g.,][]{Persson2004}, of the stars in the adopted color range. The peak of the histogram then defines the JAGB brightness for the host galaxy, $m_{\rm JAGB, F150W}=23.21\pm0.05$ mag. To determine the distance modulus, $\mu$, for NGC 1637 we scale this result to the JAGB brightness in the megamaser galaxy M106, $m_{\rm JAGB, F150W}=22.33\pm0.02$ mag \citep{Li2024,Li2025b}. The difference in brightness of the JAGB between the two galaxies therefore is 0.88 mag. If we assume the megamaser-based distance modulus for M106 is $\mu=29.397\pm0.032$ mag \citep{Reid2019}, then the distance modulus for NGC 1637 by this technique is $\mu=30.28\pm0.11$ mag, or a distance of $11.38\pm0.58$ Mpc.

A number of factors determine the uncertainty in the distance from this method. The statistical uncertainty includes an uncertainty in the mode of 0.01 mag; uncertainty due to the choice of smoothing scale of 0.04 mag (established by increasing and decreasing by 0.05 mag in the GLOESS smoothing scale); and, a ``convergence'' error of 0.02 mag, following, e.g., \citet[][essentially, the amount of variance in brightness after the first point of convergence]{Lee2025}. All together those add up in quadrature to 0.05 mag. 
The systematic uncertainty includes several individual terms, including an uncertainty in the {\tt Dolphot} aperture corrections of 0.02 mag; uncertainty in the foreground extinction of 0.01 mag; miscellaneous systematics in {\tt Dolphot} measurements (e.g., empirical point-spread-function fitting adjustments, etc.) of 0.02 mag; an average crowding correction, based on values from samples in \citet{Li2024,Li2025b}, adopted to be 0.04 mag; and an additional systematic zero-point uncertainty, composed of many individual terms, such as a geometric distance uncertainty of
0.03 mag, “methodological” variations, of how the analysis is conducted, as pointed out by the SH0ES collaboration \citep{Riess2022}, of 0.06 mag, a statistical measurement uncertainty of the JAGB feature in M106 of 0.02 mag, aperture correction uncertainties in the M106 measurement of 0.02 mag, and uncertainties due to smoothing scales of 0.04 mag.
Adding these systematic uncertainties in quadrature with the uncertainties in the JAGB magnitude results in the total 0.12 mag uncertainty in the distance modulus. 

Lastly, we caution that the JAGB technique is relatively new, and there are ongoing investigations into the potential metallicity dependencies of the JAGB feature \citep{Goldman2025}. Our uncertainty budget should be considered as a lower limit on the true uncertainty of this method.

\subsection{Host-Galaxy TRGB Distance Investigation}\label{sec:trgb}

We considered all of the available {\sl HST\/} WFPC2 data in both F555W and F814W, across all four chips of the instrument covering the host galaxy, to investigate whether it would be possible to detect the TRGB. The data from GO-9155 consisted of 12 epochs of WFPC2 F555W and 7 epochs of WFPC2 F814W images, with 1100~s exposure times for each frame. We also included the pair of 460~s exposures in F814W from GO-9042. The total depths of the CMD were ostensibly, then, a very deep 13,200~s and 8160~s in F555W and F814W, respectively. All of the frames were then run through {\tt Dolphot}. We adopted the same quality filtering scheme for these {\sl HST\/} data as we did in Section~\ref{sec:jagb} for the {\sl JWST\/} data.

We also considered the WFC3 data from GO-17502 in F555W and F814W with total exposure times 625 and 749~s, respectively. Although the depth is not nearly comparable to that afforded by the WFPC2 data, we could stand to gain with the higher WFC3/UVIS sensitivity and resolution. Again, these data were processed with {\tt Dolphot} and a similar post-processing filtering was applied on the output source detections.

\begin{figure}
\plotone{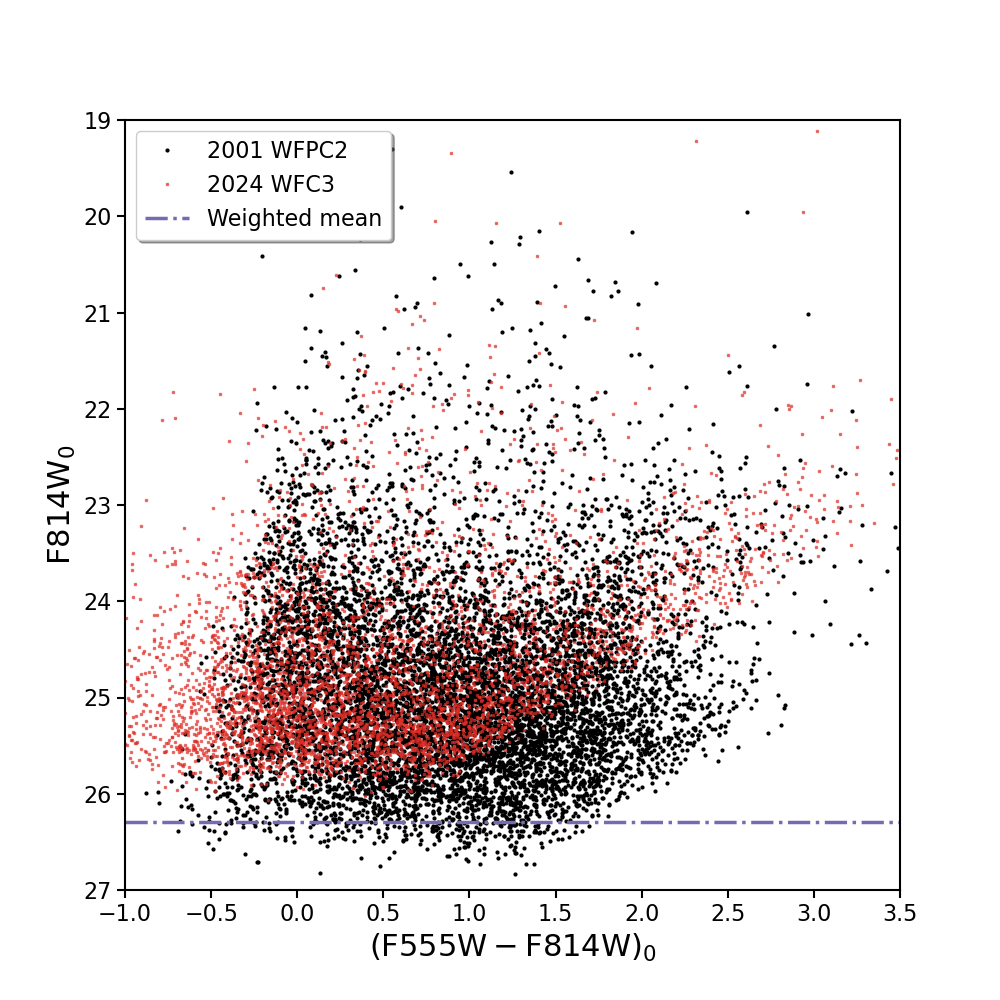}
\caption{CMD based on the WFPC2 and WFC3 UVIS data for the host galaxy NGC 1637. The {\tt Dolphot} photometry has been post-processed via quality filtering (see text) and corrected from Galactic foreground reddening. Also shown would be the brightness of the TRGB if the host distance were the adopted weighted mean of galaxy-based distance estimates (see Section~\ref{sec:distance}), assuming $M_I^0=-4.05$ mag for the TRGB \citep{Freedman2021}.
\label{fig:trgb}}
\end{figure}

We show the resulting CMD in Figure~\ref{fig:trgb}. As one can see, the WFPC2 photometry reached as deep as ${\rm F814W}_0 \sim 26.5$ mag, whereas the WFC3 observations only reach $\sim 25.8$
mag. We also show the inferred levels of the TRGB, if the host distance were the weighted mean of the galaxy-based distance estimates (see Section~\ref{sec:distance}), assuming the $I$-band absolute brightness of the TRGB is $M_I^0=-4.05$ mag \citep{Freedman2021}. As can be seen, neither the WFC3 nor the WFPC2 results are sufficiently deep to have reached the TRGB. (Of course, if the actual host distance is lower than these two estimates, then the TRGB may have been detectable, although the tip is not readily obvious in the results shown.)

\bibliography{main}{}
\bibliographystyle{aasjournalv7}



\end{document}